\def\PsfigVersion{1.10}
\def\setDriver{\DvipsDriver} 
\ifx\undefined\psfig\else\endinput\fi
\let\LaTeXAtSign=\@
\let\@=\relax
\edef\psfigRestoreAt{\catcode`\@=\number\catcode`@\relax}
\catcode`\@=11\relax
\newwrite\@unused
\def\ps@typeout#1{{\let\protect\string\immediate\write\@unused{#1}}}

\def\DvipsDriver{
	\ps@typeout{psfig/tex \PsfigVersion -dvips}
\def\PsfigSpecials{\DvipsSpecials} 	\def\ps@dir{/}
\def\ps@predir{} }
\def\OzTeXDriver{
	\ps@typeout{psfig/tex \PsfigVersion -oztex}
	\def\PsfigSpecials{\OzTeXSpecials}
	\def\ps@dir{:}
	\def\ps@predir{:}
	\catcode`\^^J=5
}

\def\figurepath{./:}

\def\DoPaths#1{\expandafter\EachPath#1\stoplist}
\def\leer{}
\def\EachPath#1:#2\stoplist{
  \ExistsFile{#1}{\SearchedFile}
  \ifx#2\leer
  \else
    \expandafter\EachPath#2\stoplist
  \fi}
\def\ps@dir{/}
\def\ExistsFile#1#2{%
   \openin1=\ps@predir#1\ps@dir#2
   \ifeof1
       \closein1
   \else
       \closein1
        \ifx\ps@founddir\leer
           \edef\ps@founddir{#1}
        \fi
   \fi}
\def\get@dir#1{%
  \def\ps@founddir{}
  \def\SearchedFile{#1}
  \DoPaths\figurepath
}

\def\@nnil{\@nil}
\def\@empty{}
\def\@psdonoop#1\@@#2#3{}
\def\@psdo#1:=#2\do#3{\edef\@psdotmp{#2}\ifx\@psdotmp\@empty \else
    \expandafter\@psdoloop#2,\@nil,\@nil\@@#1{#3}\fi}
\def\@psdoloop#1,#2,#3\@@#4#5{\def#4{#1}\ifx #4\@nnil \else
       #5\def#4{#2}\ifx #4\@nnil \else#5\@ipsdoloop #3\@@#4{#5}\fi\fi}
\def\@ipsdoloop#1,#2\@@#3#4{\def#3{#1}\ifx #3\@nnil 
       \let\@nextwhile=\@psdonoop \else
      #4\relax\let\@nextwhile=\@ipsdoloop\fi\@nextwhile#2\@@#3{#4}}
\def\@tpsdo#1:=#2\do#3{\xdef\@psdotmp{#2}\ifx\@psdotmp\@empty \else
    \@tpsdoloop#2\@nil\@nil\@@#1{#3}\fi}
\def\@tpsdoloop#1#2\@@#3#4{\def#3{#1}\ifx #3\@nnil 
       \let\@nextwhile=\@psdonoop \else
      #4\relax\let\@nextwhile=\@tpsdoloop\fi\@nextwhile#2\@@#3{#4}}
\ifx\undefined\fbox
\newdimen\fboxrule
\newdimen\fboxsep
\newdimen\ps@tempdima
\newbox\ps@tempboxa
\long\def\fbox#1{\leavevmode\setbox\ps@tempboxa\hbox{#1}\ps@tempdima\fboxrule
    \advance\ps@tempdima \fboxsep \advance\ps@tempdima \dp\ps@tempboxa
   \hbox{\lower \ps@tempdima\hbox
  {\vbox{\hrule height \fboxrule
          \hbox{\vrule width \fboxrule \hskip\fboxsep
          \vbox{\vskip\fboxsep \box\ps@tempboxa\vskip\fboxsep}\hskip 
                 \fboxsep\vrule width \fboxrule}
                 \hrule height \fboxrule}}}}
\fi
\newread\ps@stream
\newif\ifnot@eof       
\newif\if@noisy        
\newif\if@atend        
\newif\if@psfile       
{\catcode`\%=12\global\gdef\epsf@start{
\def\epsf@PS{PS}
\def\epsf@getbb#1{%
\openin\ps@stream=\ps@predir#1
\ifeof\ps@stream\ps@typeout{Error, File #1 not found}\else
   {\not@eoftrue \chardef\other=12
    \def\do##1{\catcode`##1=\other}\dospecials \catcode`\ =10
    \loop
       \if@psfile
	  \read\ps@stream to \epsf@fileline
       \else{
	  \obeyspaces
          \read\ps@stream to \epsf@tmp\global\let\epsf@fileline\epsf@tmp}
       \fi
       \ifeof\ps@stream\not@eoffalse\else
       \if@psfile\else
       \expandafter\epsf@test\epsf@fileline:. \\%
       \fi
          \expandafter\epsf@aux\epsf@fileline:. \\%
       \fi
   \ifnot@eof\repeat
   }\closein\ps@stream\fi}%
\long\def\epsf@test#1#2#3:#4\\{\def\epsf@testit{#1#2}
			\ifx\epsf@testit\epsf@start\else
\ps@typeout{Warning! File does not start with `\epsf@start'.  It may not be a PostScript file.}
			\fi
			\@psfiletrue} 
{\catcode`\%=12\global\let\epsf@percent=
\long\def\epsf@aux#1#2:#3\\{\ifx#1\epsf@percent
   \def\epsf@testit{#2}\ifx\epsf@testit\epsf@bblit
	\@atendfalse
        \epsf@atend #3 . \\%
	\if@atend	
	   \if@verbose{
		\ps@typeout{psfig: found `(atend)'; continuing search}
	   }\fi
        \else
        \epsf@grab #3 . . . \\%
        \not@eoffalse
        \global\no@bbfalse
        \fi
   \fi\fi}%
\def\epsf@grab #1 #2 #3 #4 #5\\{%
   \global\def\epsf@llx{#1}\ifx\epsf@llx\empty
      \epsf@grab #2 #3 #4 #5 .\\\else
   \global\def\epsf@lly{#2}%
   \global\def\epsf@urx{#3}\global\def\epsf@ury{#4}\fi}%
\def\epsf@atendlit{(atend)} 
\def\epsf@atend #1 #2 #3\\{%
   \def\epsf@tmp{#1}\ifx\epsf@tmp\empty
      \epsf@atend #2 #3 .\\\else
   \ifx\epsf@tmp\epsf@atendlit\@atendtrue\fi\fi}

\chardef\psletter = 11 
\chardef\other = 12

\newif \ifdebug 
\newif\ifc@mpute 
\c@mputetrue 

\let\then = \relax
\def\r@dian{pt }
\let\r@dians = \r@dian
\let\dimensionless@nit = \r@dian
\let\dimensionless@nits = \dimensionless@nit
\def\internal@nit{sp }
\let\internal@nits = \internal@nit
\newif\ifstillc@nverging
\def \Mess@ge #1{\ifdebug \then \message {#1} \fi}

{ 
	\catcode `\@ = \psletter
	\gdef \nodimen {\expandafter \n@dimen \the \dimen}
	\gdef \term #1 #2 #3%
	       {\edef \t@ {\the #1}
		\edef \t@@ {\expandafter \n@dimen \the #2\r@dian}%
		\t@rm {\t@} {\t@@} {#3}%
	       }
	\gdef \t@rm #1 #2 #3%
	       {{%
		\count 0 = 0
		\dimen 0 = 1 \dimensionless@nit
		\dimen 2 = #2\relax
		\Mess@ge {Calculating term #1 of \nodimen 2}%
		\loop
		\ifnum	\count 0 < #1
		\then	\advance \count 0 by 1
			\Mess@ge {Iteration \the \count 0 \space}%
			\Multiply \dimen 0 by {\dimen 2}%
			\Mess@ge {After multiplication, term = \nodimen 0}%
			\Divide \dimen 0 by {\count 0}%
			\Mess@ge {After division, term = \nodimen 0}%
		\repeat
		\Mess@ge {Final value for term #1 of 
				\nodimen 2 \space is \nodimen 0}%
		\xdef \Term {#3 = \nodimen 0 \r@dians}%
		\aftergroup \Term
	       }}
	\catcode `\p = \other
	\catcode `\t = \other
	\gdef \n@dimen #1pt{#1} 
}

\def \Divide #1by #2{\divide #1 by #2} 

\def \Multiply #1by #2
       {{
	\count 0 = #1\relax
	\count 2 = #2\relax
	\count 4 = 65536
	\Mess@ge {Before scaling, count 0 = \the \count 0 \space and
			count 2 = \the \count 2}%
	\ifnum	\count 0 > 32767 
	\then	\divide \count 0 by 4
		\divide \count 4 by 4
	\else	\ifnum	\count 0 < -32767
		\then	\divide \count 0 by 4
			\divide \count 4 by 4
		\else
		\fi
	\fi
	\ifnum	\count 2 > 32767 
	\then	\divide \count 2 by 4
		\divide \count 4 by 4
	\else	\ifnum	\count 2 < -32767
		\then	\divide \count 2 by 4
			\divide \count 4 by 4
		\else
		\fi
	\fi
	\multiply \count 0 by \count 2
	\divide \count 0 by \count 4
	\xdef \product {#1 = \the \count 0 \internal@nits}%
	\aftergroup \product
       }}

\def\r@duce{\ifdim\dimen0 > 90\r@dian \then   
		\multiply\dimen0 by -1
		\advance\dimen0 by 180\r@dian
		\r@duce
	    \else \ifdim\dimen0 < -90\r@dian \then  
		\advance\dimen0 by 360\r@dian
		\r@duce
		\fi
	    \fi}

\def\Sine#1%
       {{%
	\dimen 0 = #1 \r@dian
	\r@duce
	\ifdim\dimen0 = -90\r@dian \then
	   \dimen4 = -1\r@dian
	   \c@mputefalse
	\fi
	\ifdim\dimen0 = 90\r@dian \then
	   \dimen4 = 1\r@dian
	   \c@mputefalse
	\fi
	\ifdim\dimen0 = 0\r@dian \then
	   \dimen4 = 0\r@dian
	   \c@mputefalse
	\fi
	\ifc@mpute \then
		\divide\dimen0 by 180
		\dimen0=3.141592654\dimen0
		\dimen 2 = 3.1415926535897963\r@dian 
		\divide\dimen 2 by 2 
		\Mess@ge {Sin: calculating Sin of \nodimen 0}%
		\count 0 = 1 
		\dimen 2 = 1 \r@dian 
		\dimen 4 = 0 \r@dian 
		\loop
			\ifnum	\dimen 2 = 0 
			\then	\stillc@nvergingfalse 
			\else	\stillc@nvergingtrue
			\fi
			\ifstillc@nverging 
			\then	\term {\count 0} {\dimen 0} {\dimen 2}%
				\advance \count 0 by 2
				\count 2 = \count 0
				\divide \count 2 by 2
				\ifodd	\count 2 
				\then	\advance \dimen 4 by \dimen 2
				\else	\advance \dimen 4 by -\dimen 2
				\fi
		\repeat
	\fi		
			\xdef \sine {\nodimen 4}%
       }}

\def\Cosine#1{\ifx\sine\UnDefined\edef\Savesine{\relax}\else
		             \edef\Savesine{\sine}\fi
	{\dimen0=#1\r@dian\advance\dimen0 by 90\r@dian
	 \Sine{\nodimen 0}
	 \xdef\cosine{\sine}
	 \xdef\sine{\Savesine}}}	      

\def\psdraft{
	\def\@psdraft{0}
}
\def\psfull{
	\def\@psdraft{100}
}

\psfull  

\newif\if@scalefirst
\def\psscalefirst{\@scalefirsttrue}
\def\psrotatefirst{\@scalefirstfalse}
\psrotatefirst

\newif\if@draftbox
\def\psnodraftbox{
	\@draftboxfalse
}
\def\psdraftbox{
	\@draftboxtrue
}
\@draftboxtrue

\newif\if@prologfile
\newif\if@postlogfile
\def\pssilent{
	\@noisyfalse
}
\def\psnoisy{
	\@noisytrue
}
\psnoisy
\newif\if@bbllx
\newif\if@bblly
\newif\if@bburx
\newif\if@bbury
\newif\if@height
\newif\if@width
\newif\if@rheight
\newif\if@rwidth
\newif\if@angle
\newif\if@clip
\newif\if@verbose
\def\@p@@sclip#1{\@cliptrue}
\newif\if@decmpr
\def\@p@@sfigure#1{\def\@p@sfile{null}\def\@p@sbbfile{null}\@decmprfalse
   \openin1=\ps@predir#1
   \ifeof1
	\closein1
	\get@dir{#1}
	\ifx\ps@founddir\leer
		\openin1=\ps@predir#1.bb
		\ifeof1
			\closein1
			\get@dir{#1.bb}
			\ifx\ps@founddir\leer
				\ps@typeout{Can't find #1 in \figurepath}
			\else
				\@decmprtrue
				\def\@p@sfile{\ps@founddir\ps@dir#1}
				\def\@p@sbbfile{\ps@founddir\ps@dir#1.bb}
			\fi
		\else
			\closein1
			\@decmprtrue
			\def\@p@sfile{#1}
			\def\@p@sbbfile{#1.bb}
		\fi
	\else
		\def\@p@sfile{\ps@founddir\ps@dir#1}
		\def\@p@sbbfile{\ps@founddir\ps@dir#1}
	\fi
   \else
	\closein1
	\def\@p@sfile{#1}
	\def\@p@sbbfile{#1}
   \fi
}
\def\@p@@sfile#1{\@p@@sfigure{#1}}
\def\@p@@sbbllx#1{
		\@bbllxtrue
		\dimen100=#1
		\edef\@p@sbbllx{\number\dimen100}
}
\def\@p@@sbblly#1{
		\@bbllytrue
		\dimen100=#1
		\edef\@p@sbblly{\number\dimen100}
}
\def\@p@@sbburx#1{
		\@bburxtrue
		\dimen100=#1
		\edef\@p@sbburx{\number\dimen100}
}
\def\@p@@sbbury#1{
		\@bburytrue
		\dimen100=#1
		\edef\@p@sbbury{\number\dimen100}
}
\def\@p@@sheight#1{
		\@heighttrue
		\dimen100=#1
   		\edef\@p@sheight{\number\dimen100}
}
\def\@p@@swidth#1{
		\@widthtrue
		\dimen100=#1
		\edef\@p@swidth{\number\dimen100}
}
\def\@p@@srheight#1{
		\@rheighttrue
		\dimen100=#1
		\edef\@p@srheight{\number\dimen100}
}
\def\@p@@srwidth#1{
		\@rwidthtrue
		\dimen100=#1
		\edef\@p@srwidth{\number\dimen100}
}
\def\@p@@sangle#1{
		\@angletrue
		\edef\@p@sangle{#1} 
}
\def\@p@@ssilent#1{ 
		\@verbosefalse
}
\def\@p@@sprolog#1{\@prologfiletrue\def\@prologfileval{#1}}
\def\@p@@spostlog#1{\@postlogfiletrue\def\@postlogfileval{#1}}
\def\@cs@name#1{\csname #1\endcsname}
\def\@setparms#1=#2,{\@cs@name{@p@@s#1}{#2}}
\def\ps@init@parms{
		\@bbllxfalse \@bbllyfalse
		\@bburxfalse \@bburyfalse
		\@heightfalse \@widthfalse
		\@rheightfalse \@rwidthfalse
		\def\@p@sbbllx{}\def\@p@sbblly{}
		\def\@p@sbburx{}\def\@p@sbbury{}
		\def\@p@sheight{}\def\@p@swidth{}
		\def\@p@srheight{}\def\@p@srwidth{}
		\def\@p@sangle{0}
		\def\@p@sfile{} \def\@p@sbbfile{}
		\def\@p@scost{10}
		\def\@sc{}
		\@prologfilefalse
		\@postlogfilefalse
		\@clipfalse
		\if@noisy
			\@verbosetrue
		\else
			\@verbosefalse
		\fi
}
\def\parse@ps@parms#1{
	 	\@psdo\@psfiga:=#1\do
		   {\expandafter\@setparms\@psfiga,}}
\newif\ifno@bb
\def\bb@missing{
	\if@verbose{
		\ps@typeout{psfig: searching \@p@sbbfile \space  for bounding box}
	}\fi
	\no@bbtrue
	\epsf@getbb{\@p@sbbfile}
        \ifno@bb \else \bb@cull\epsf@llx\epsf@lly\epsf@urx\epsf@ury\fi
}	
\def\bb@cull#1#2#3#4{
	\dimen100=#1 bp\edef\@p@sbbllx{\number\dimen100}
	\dimen100=#2 bp\edef\@p@sbblly{\number\dimen100}
	\dimen100=#3 bp\edef\@p@sbburx{\number\dimen100}
	\dimen100=#4 bp\edef\@p@sbbury{\number\dimen100}
	\no@bbfalse
}
\newdimen\p@intvaluex
\newdimen\p@intvaluey
\def\rotate@#1#2{{\dimen0=#1 sp\dimen1=#2 sp
		  \global\p@intvaluex=\cosine\dimen0
		  \dimen3=\sine\dimen1
		  \global\advance\p@intvaluex by -\dimen3
		  \global\p@intvaluey=\sine\dimen0
		  \dimen3=\cosine\dimen1
		  \global\advance\p@intvaluey by \dimen3
		  }}
\def\compute@bb{
		\no@bbfalse
		\if@bbllx \else \no@bbtrue \fi
		\if@bblly \else \no@bbtrue \fi
		\if@bburx \else \no@bbtrue \fi
		\if@bbury \else \no@bbtrue \fi
		\ifno@bb \bb@missing \fi
		\ifno@bb \ps@typeout{FATAL ERROR: no bb supplied or found}
			\no-bb-error
		\fi
		\count203=\@p@sbburx
		\count204=\@p@sbbury
		\advance\count203 by -\@p@sbbllx
		\advance\count204 by -\@p@sbblly
		\edef\ps@bbw{\number\count203}
		\edef\ps@bbh{\number\count204}
		\if@angle 
			\Sine{\@p@sangle}\Cosine{\@p@sangle}
	        	{\dimen100=\maxdimen\xdef\r@p@sbbllx{\number\dimen100}
					    \xdef\r@p@sbblly{\number\dimen100}
			                    \xdef\r@p@sbburx{-\number\dimen100}
					    \xdef\r@p@sbbury{-\number\dimen100}}
                        \def\minmaxtest{
			   \ifnum\number\p@intvaluex<\r@p@sbbllx
			      \xdef\r@p@sbbllx{\number\p@intvaluex}\fi
			   \ifnum\number\p@intvaluex>\r@p@sbburx
			      \xdef\r@p@sbburx{\number\p@intvaluex}\fi
			   \ifnum\number\p@intvaluey<\r@p@sbblly
			      \xdef\r@p@sbblly{\number\p@intvaluey}\fi
			   \ifnum\number\p@intvaluey>\r@p@sbbury
			      \xdef\r@p@sbbury{\number\p@intvaluey}\fi
			   }
			\rotate@{\@p@sbbllx}{\@p@sbblly}
			\minmaxtest
			\rotate@{\@p@sbbllx}{\@p@sbbury}
			\minmaxtest
			\rotate@{\@p@sbburx}{\@p@sbblly}
			\minmaxtest
			\rotate@{\@p@sbburx}{\@p@sbbury}
			\minmaxtest
			\edef\@p@sbbllx{\r@p@sbbllx}\edef\@p@sbblly{\r@p@sbblly}
			\edef\@p@sbburx{\r@p@sbburx}\edef\@p@sbbury{\r@p@sbbury}
		\fi
		\count203=\@p@sbburx
		\count204=\@p@sbbury
		\advance\count203 by -\@p@sbbllx
		\advance\count204 by -\@p@sbblly
		\edef\@bbw{\number\count203}
		\edef\@bbh{\number\count204}
}
\def\in@hundreds#1#2#3{\count240=#2 \count241=#3
		     \count100=\count240	
		     \divide\count100 by \count241
		     \count101=\count100
		     \multiply\count101 by \count241
		     \advance\count240 by -\count101
		     \multiply\count240 by 10
		     \count101=\count240	
		     \divide\count101 by \count241
		     \count102=\count101
		     \multiply\count102 by \count241
		     \advance\count240 by -\count102
		     \multiply\count240 by 10
		     \count102=\count240	
		     \divide\count102 by \count241
		     \count200=#1\count205=0
		     \count201=\count200
			\multiply\count201 by \count100
		 	\advance\count205 by \count201
		     \count201=\count200
			\divide\count201 by 10
			\multiply\count201 by \count101
			\advance\count205 by \count201
		     \count201=\count200
			\divide\count201 by 100
			\multiply\count201 by \count102
			\advance\count205 by \count201
		     \edef\@result{\number\count205}
}
\def\compute@wfromh{
		\in@hundreds{\@p@sheight}{\@bbw}{\@bbh}
		\edef\@p@swidth{\@result}
}
\def\compute@hfromw{
	        \in@hundreds{\@p@swidth}{\@bbh}{\@bbw}
		\edef\@p@sheight{\@result}
}
\def\compute@handw{
		\if@height 
			\if@width
			\else
				\compute@wfromh
			\fi
		\else 
			\if@width
				\compute@hfromw
			\else
				\edef\@p@sheight{\@bbh}
				\edef\@p@swidth{\@bbw}
			\fi
		\fi
}
\def\compute@resv{
		\if@rheight \else \edef\@p@srheight{\@p@sheight} \fi
		\if@rwidth \else \edef\@p@srwidth{\@p@swidth} \fi
}
\def\compute@sizes{
	\compute@bb
	\if@scalefirst\if@angle
	\if@width
	   \in@hundreds{\@p@swidth}{\@bbw}{\ps@bbw}
	   \edef\@p@swidth{\@result}
	\fi
	\if@height
	   \in@hundreds{\@p@sheight}{\@bbh}{\ps@bbh}
	   \edef\@p@sheight{\@result}
	\fi
	\fi\fi
	\compute@handw
	\compute@resv}
\def\OzTeXSpecials{
	\special{empty.ps /@isp {true} def}
	\special{empty.ps \@p@swidth \space \@p@sheight \space
			\@p@sbbllx \space \@p@sbblly \space
			\@p@sbburx \space \@p@sbbury \space
			startTexFig \space }
	\if@clip{
		\if@verbose{
			\ps@typeout{(clip)}
		}\fi
		\special{empty.ps doclip \space }
	}\fi
	\if@angle{
		\if@verbose{
			\ps@typeout{(rotate)}
		}\fi
		\special {empty.ps \@p@sangle \space rotate \space} 
	}\fi
	\if@prologfile
	    \special{\@prologfileval \space } \fi
	\if@decmpr{
		\if@verbose{
			\ps@typeout{psfig: Compression not available
			in OzTeX version \space }
		}\fi
	}\else{
		\if@verbose{
			\ps@typeout{psfig: including \@p@sfile \space }
		}\fi
		\special{epsf=\ps@predir\@p@sfile \space }
	}\fi
	\if@postlogfile
	    \special{\@postlogfileval \space } \fi
	\special{empty.ps /@isp {false} def}
}
\def\DvipsSpecials{
	\special{ps::[begin] 	\@p@swidth \space \@p@sheight \space
			\@p@sbbllx \space \@p@sbblly \space
			\@p@sbburx \space \@p@sbbury \space
			startTexFig \space }
	\if@clip{
		\if@verbose{
			\ps@typeout{(clip)}
		}\fi
		\special{ps:: doclip \space }
	}\fi
	\if@angle
		\if@verbose{
			\ps@typeout{(clip)}
		}\fi
		\special {ps:: \@p@sangle \space rotate \space} 
	\fi
	\if@prologfile
	    \special{ps: plotfile \@prologfileval \space } \fi
	\if@decmpr{
		\if@verbose{
			\ps@typeout{psfig: including \@p@sfile.Z \space }
		}\fi
		\special{ps: plotfile "`zcat \@p@sfile.Z" \space }
	}\else{
		\if@verbose{
			\ps@typeout{psfig: including \@p@sfile \space }
		}\fi
		\special{ps: plotfile \@p@sfile \space }
	}\fi
	\if@postlogfile
	    \special{ps: plotfile \@postlogfileval \space } \fi
	\special{ps::[end] endTexFig \space }
}
\def\psfig#1{\vbox {
	\ps@init@parms
	\parse@ps@parms{#1}
	\compute@sizes
	\ifnum\@p@scost<\@psdraft{
		\PsfigSpecials 
		\vbox to \@p@srheight sp{
			\hbox to \@p@srwidth sp{
				\hss
			}
		\vss
		}
	}\else{
		\if@draftbox{		
			\hbox{\fbox{\vbox to \@p@srheight sp{
			\vss
			\hbox to \@p@srwidth sp{ \hss 
			 \hss }
			\vss
			}}}
		}\else{
			\vbox to \@p@srheight sp{
			\vss
			\hbox to \@p@srwidth sp{\hss}
			\vss
			}
		}\fi

	}\fi
}}
\psfigRestoreAt
\setDriver
\let\@=\LaTeXAtSign

 \documentstyle[11pt,aaspp]{article}
\def \m{\ifmmode M_\odot\else M$_\odot$\fi}
\def\e#1{$\times$ 10$^{#1}$}
\def\etal{{\it et al.}}
\def\ni{$^{56}$Ni}
\def \lta {\mathrel{\vcenter 
     {\hbox{$<$}\nointerlineskip\hbox{$\sim$}}}} 
\def \gta {\mathrel{\vcenter 
     {\hbox{$>$}\nointerlineskip\hbox{$\sim$}}}} 
\tighten

\def\deg{\ifmmode^\circ\else$^\circ$\fi} 

\begin{document}

\title{ANALYSIS OF THE POLARIZATION AND FLUX SPECTRA OF SN1993J}

\author{P. H\"oflich}          
\affil{Center for Astrophysics, Harvard University}
\authoraddr{ Cambridge,  MA 02138, U.S.A.}

\author{J. C. Wheeler} 
\affil{Department of Astronomy} 
\authoraddr{University of Texas, Austin, TX 78712}

\author{D. C. Hines} 
\affil{Steward Observatory} 
\authoraddr{Steward Observatory, 933 N. Cherry Ave. Tucson, AZ 85721}
  
\author{S. R. Trammell} 
\affil{The Enrico Fermi Institute}
\authoraddr{The University of Chicago, 5640 South Ellis Ave., \\
Chicago, IL 60637}

\noindent
{\bf ABSTRACT:}
Synthetic polarization and flux spectra are presented for aspherical, 
electron scattering-dominated photospheres of Type II supernovae (SN~II) in 
general and the specific case of SN 1993J.  Monte Carlo calculations are 
based on the following assumptions: a) ellipsoidal envelopes with power law 
density profiles; b) occupation numbers given by local thermodynamical 
equilibrium (LTE); c) pure electron scattering for continuum opacities; d) 
lines treated in a Sobolev approximation with an assumed constant
thermalization fraction; e) 
line transitions result in depolarization; and f) the temperature structure 
is given by a gray extended atmosphere.
 
The observed luminosity of a Type II supernova depends on the unknown 
inclination angle $i$.  Spectral analysis 
alone will fail to detect even strong deviations from spherical symmetry.  
Line scattering 
depolarizes incident polarized light, but the residual intrinsic 
polarization does not completely vanish because of electron scattering 
effects that depend on the electron density distribution.  By combining 
results on the polarization and velocity structure of the emission lines, 
and the degree of polarization in the continuum, we place strong constraints 
on the degree of asphericity, the inclination of the system and the 
electron density distribution. In addition, since the problem is over 
constrained, we can independently test for the contribution to the 
polarization caused by aligned interstellar grains between 
the SN and earth. 

These modelling techniques have been applied to SN1993J. Both the 
flux spectra and the percentage of polarization as a function
of wavelength can be reproduced by an aspherical model with 
an axis ratio of 0.6, radial electron densities proportional to $r^{-5}$ 
and an effective temperature of 4800 K.  In an oblate model, SN 1993J is 
seen almost equator on.  The line forming region is still within the 
hydrogen rich part of the envelope 3 weeks after the explosion, although 
the continuum may form in deeper layers.  Although the 
residual polarization across H$\alpha$ is not zero, the interstellar 
component derived herein is consistent with that deduced by Trammell \etal\
(1993).  
The power of this technique for investigating the  structure of other SN~II,
as a method for
independently deriving the  interstellar polarization, and the implications
of these results 
on the use of SN II to determine distances through the
Baade-Wesselink  method are discussed.

{\sl Keywords}:{Radiative transfer --- polarization --- supernovae: general --- supernovae: individual (SN1993J) 
}                
\vfill
 
\eject 
 
\section{Introduction} 
 
The investigation of the emitted light of Type II supernovae (SN~II) is 
important for various fields in astronomy and astrophysics.  The observed 
spectra, spectropolarimetry and light curves give direct information on 
the physical, geometrical and chemical conditions of the expanding 
envelope.  In principle, this allows us to test dynamic models of the 
ejecta from supernovae, and investigate the explosion mechanism of SN~II 
and the final stages of the evolution of massive stars.  Because SN~II are 
among the brightest single objects they are also commonly used as distance 
indicators based on the Baade-Wesselink  method 
(Baade 1926, Wesselink 1946, H\"oflich \etal\ 1986, H\"oflich 1988, 1991,
Eastman \& Kirshner 1989,  
Baron \etal\ 1994, Schmidt \etal\ 1994), 
but this method     
depends critically on the assumption of spherical envelopes.  Little 
attention has been paid, however, to polarization measurements that are 
crucial to test this assumption and investigate the overall geometry.

Very few polarization measurements are available for supernovae.
SN 1968L (Serkowski 1970) and SN 1970G (Shakhovskoi and Efimov 1973)
were Type II. SN 1968L was not isolated from the ISM but 
SN 1970G was probably intrinsically polarized at the level
of 0.5 -- 1 percent. Three Type Ia events have been reported with no detectable
polarization above the interstellar component, SN 1972E
(Wolstencraft and Kemp 1972), SN 1983G (McCall
\etal 1984; McCall 1985), and SN 1992A (Spyromilio and Bailey 1993).
One Type Ib, SN 1983N, known to have a significant
circumstellar nebula, has been detected (McCall \etal\ 1984;
McCall 1985).  
Several recent events have also been observed, some Type Ia with
low polarization (SN 1994D, SN 1994ae, and SN 1995D), 
SN 1994Y, a narrow-line Type II that shows polarization at
the level of 1.5 -- 2 percent and SN 1995H, another SN II
with detectable polarization at about the 1 percent level (Wang \etal\ 1995).  

The best polarization data to date are for
SN 1987A and SN1993J. 
For SN 1987A, polarization of up to 
0.5\% was observed early on, but decreased rapidly over the next few weeks 
(Mendez \etal\ 1988; Clocchiatti \etal\ 1988; Cropper \etal\ 1988).  
It was concluded that the envelope of SN 1987A showed 
asphericities of about 10 to 20\% and that Thomson scattering was the main 
physical effect causing the polarization
%
%
 (e.g. Mendez et al. 1988, H\"oflich 1991 and Jeffery 1991).  Whether 
the asymmetry implied by the polarization measurements is caused by rapid 
rotation of the progenitor (H\"oflich \& Steinmetz 1991, Steinmetz \& 
H\"oflich 1992) or by an
asphericity of the  SN~II-explosion mechanism itself (Yamada \& Sato 1991)
is still a 
question under debate.  Depending on the answer, asphericity, and thus 
observable polarization, may prove to be a common property of SN~II or just 
related to the special nature of SN 1987A 
(Arnett \etal\ 1989, Hillebrandt \& H\"oflich 1990).  If it is common,
then it may significantly impact the use of supernova to estimate
distances (McCall 1985).
 
SN 1993J was observed from the beginning in wavelength bands from X-ray to 
radio.  Soon after the discovery the spectral evolution and the light 
curves made it clear that SN1993J -- like SN 1987A -- did not represent a 
typical SN~II.  Although details related to the progenitor evolution and 
the relation between initial mass and He/C-core mass are disputed, general 
agreement was soon achieved that SN1993J represents an object that 
had lost most of its H-rich envelope prior to the explosion with a 
remaining H-rich mass of $\approx 0.2 -- 1.5$ \m\ (H\"oflich, 
Langer \& Duschinger, 1993; Wheeler \etal\ 1993, Swartz \etal\ 1993; 
Woosley \etal 1994; Shigeyama \etal\ 1994; Jeffery \etal\ 1994). 
The progenitor was a red or yellow supergiant that underwent very 
strong mass loss, as confirmed by observations in the radio and X-ray
(Van Dyk \etal\ 1994; Fransson, Lundqvist \& Chevalier, 1994).  
For reviews see Wheeler \& Filippenko (1995) and
Baron \etal\ (1995).
Unlike SN 1987A, the polarization was already high ($\gta$ 1 percent) 
in SN 1993J only a few days after the onset of explosion
and increased to $\approx$ 1.6 percent 3 weeks later (see 
\S\ 4).  A strongly differential stellar rotation, as 
suggested for SN 1987A, can be ruled out because the progenitor of
SN1993J had an extended and hence presumably slowly rotating envelope.
H\"oflich (1995) studied different configurations which may explain
the polarization in the continuum.  From the size of the polarization
and its weak dependence on time, H\"oflich concluded that an
aspherical explosion is unlikely.  Neither mechanism suggested for SN
1987A is thus easily applicable in this case.  H\"oflich found that
the polarization can be explained by an oblate ellipsoid with axis
ratio of 0.6 or by ``off-center" sources which may have been caused by
the explosion in a very close binary system or by a strong ``kick" of
the neutron star during the explosion.  In principle, these scenarios
can be distinguished by the time evolution of the polarization, but a
lack of observations has prevented clarification.  A critical point is
that the value of 0.6 for the asphericity is a lower limit, valid only
if SN 1993J is seen from the equatorial plane.  H\"oflich's analysis
relied on the correction of the measured total polarization for the
interstellar component (Trammell, Hines \& Wheeler 1993).  As the
polarization is a (pseudo)vector, the size and temporal behavior of
the intrinsic component of the polarization depend sensitively on the
correction for the interstellar component.  An independent check of
the value of the ISM polarization derived by Trammell \etal\ (1993) is
appropriate.
 
Asphericity may be a common phenomenon in SN~II since it has been observed 
in both cases which have been sufficiently bright to allow for polarization 
measurements.  In this paper we will study 
asphericity effects both on the total flux spectra  
and the percent polarization as a function of wavelength.
In this paper these will be referred to as the {\it flux
spectra} and {\it polarization spectra}, respectively.  Note that
the percent polarization is to be distinguished from the 
Stokes flux spectrum as constructed and presented in Trammell \etal\
(their Fig. 2; see also \S\ 4). 
The following questions shall be addressed: Does the 
wavelength dependence of the polarization provide additional information on 
the envelope structure? Can we determine the inclination $i$ at which a 
supernovae is observed? Can the interstellar component be determined 
independently? Do the flux spectra constrain possible asphericity effects 
and can we infer from successful fits of SN~II spectra with spherical 
models that the envelopes are indeed spherical, as suggested by Baron 
\etal\ (1994).
 
The underlying physical model is described in \S\ 2.  General results of 
our calculations are discussed in the following section.  Polarimetric 
observations of SN 1993J are summarized in \S\ 4 and a comparison of the 
current model with observations is given in \S\ 5.  Implications for the 
density and chemical structure of SN 1993J are discussed.  Consistency 
checks between flux and polarization spectra are performed to test the 
interstellar polarization correction applied by Trammell \etal\ (1993).  
Our conclusions are presented in \S\ 6.
 
\section{Computational Method}
 
The current calculations use a modified 
version of the Monte-Carlo code previously applied to 
calculate the continuum polarization in SN 1987A and SN 1993J (H\"oflich 
1988, 1991, 1995).  The code is capable of handling arbitrary 
3-dimensional geometries, both for the density and the distribution of the 
sources.  Polarization and flux spectra for rapidly expanding envelopes can 
thus be computed.

Little is known about the shape of the envelopes in SN~II in general and 
SN 1993J in particular.  Therefore, we use oblate ellipsoids parameterized 
by
\begin{equation}
    x^2 + y^2 + z^2/E^2  = r^2 ,      
\end{equation}
where $x$, $y$, $z$ are orthonormal directions, E is the axis ratio of a 
rotational ellipsoid and $r \epsilon [A_{min},A_{max}]$.  $A_{min,max}$ are 
the minimum and maximum distances in the equatorial plane of the inner and
outer boundaries, respectively. 
The inner boundary is at $\tau_{sc}=3$ (the approximate thermalization
depth, see below).
H\"oflich (1995) found that the continuum 
polarization observed for SN1993J can be understood in terms of ellipsoids.  
In the following E = 0.6 is used as a canonical value.

Shortly (hours to a few days) after the core collapse, the expansion of the 
shell becomes homologous, i.e., the velocity becomes proportional to the
distance
from the center.
Piecewise, the density slope can be approximated by power laws, i.e.  $\rho 
\propto r^{-n}$. The ratio of density scale height 
to radius is thus $1/n$.  

For computations of polarization and flux spectra, 
line blocking is included in the 
Sobolev approximation (Sobolev 1957).  The implementation is similar to that 
described by Abbot and Lucy (1985) for their one dimensional Monte-Carlo code.
 
It is well known that deviations from local thermodynamical equilibrium are 
important if details of individual lines are to be understood (see e.g.  
Lucy 1988, H\"oflich 1988, Eastman \& Kirshner 1989, Baron \etal\ 1994).
Detailed NLTE 
calculations consistent with 3-D radiation transport including polarization 
are beyond the capabilities of available computers.  On the other hand, at 
least qualitative agreement with the observations can be achieved if 
LTE-population numbers are assumed 
(Branch \etal\ 1981, Harkness 1991, Jeffery \etal\ 1994, Wheeler \etal\ 1994).  
Several 
simplifications are thus adopted similar to those applied in the framework of 
spectral modelling in 1-D calculations.
The population numbers are assumed to be in local thermodynamic
equilibrium.  For He, the importance of non-thermal excitation by gamma 
rays is well established (Swartz \etal\ 1993).
This effect is mimicked here by an artificial 
enhancement of the He opacity (see discussion in \S 5). 

Thomson scattering is assumed to be the only source of opacity in the
continuum.  Lines are treated in a scattering approximation with the
thermalization fraction $\varepsilon$ of $10^{-2}$.  This value is
based on detailed NLTE-calculations (H\"oflich 1995).  If a photon is
absorbed in a line, it will be re-emitted at the same frequency (in
the comoving frame) with a probability of ($1 - \varepsilon$) or it
will be thermalized, i.e.  re-emitted at a frequency randomly chosen
according to the local black body function.  In general, complete
redistribution can be assumed for the line profiles (Mihalas 1978)
since collisional processes dominate the relative level populations
within the fine structure of the upper and lower atomic level of a
line transition.  Time scales for collisional redistribution are much
shorter than absorption and re-emission processes in a line.  Since
collisions are random processes, all information on non-local
processes (i.e.  polarization) is lost.  Therefore, we can assume that
a photon becomes unpolarized if absorbed and re-emitted in a line.

The temperature structure T(r) 
is given by an extended gray atmosphere calculation 
which gives a good approximation to NLTE atmospheres 
Eastman and Kirchner 1989; H\"oflich 1989, 1990, 1991b).  At 
small optical depths ($\tau \leq 0.3$) the excitation temperature
is taken to be constant since the lines are mainly 
scattering dominated.  

The thermalization optical depth
of continuum photons is about 3--5 Thomson scattering optical depths (e.g.  
Hershkowitz, Lindner \& Wagoner, 1986; H\"oflich \etal\ 1986) where $ 
\tau $ is measured in the radial direction.  The light becomes unpolarized 
at large optical depths, both by thermalization and multiple scattering 
processes.  The inner boundary condition is given by isotropic emission at 
the thermalization depth $\tau_{max}$.  Photons are absorbed and re-emitted 
if they propagate to the inner boundary during their random walk.  In 
contrast to the optically thin case (Brown \& McLean 1977) the 
polarization depends only slightly on $\tau_{max} $ for $\tau_{max} \gta 2$ 
(Daniel 1982; H\"oflich 1991).  Thus, we use $\tau_{max}=3 $ as 
our standard-value.  Accordingly, at this inner boundary, the radiation 
field is assumed to be given by a local black body.

The optical depth is closely related to 
the question of the photospheric radius
which is often used as a basic parameter to characterize 
supernovae atmospheres.                          
This radius is that which would be observed in the spectral 
continuum if the object were resolved in angle.  It may be defined by the 
radial distance $R_{true}$ where the optical depth in true absorption 
becomes unity, $R_{therm}$ where the photons thermalize (i.e.  the radius 
at which most of the emitted photons are formed), or $R_{sc}$ where the 
photons most likely undergo electron scattering before escape.  In 
addition, the photospheric radius is a function of frequency and
that function will vary with the definition of opacity, 
namely whether monochromatic, broad band, Rosseland 
or Planck means are used.  For a more detailed discussion see e.g.  
H\"oflich (1987, 1990, 1995; Baschek, Scholz, and Wehrse 1991) 
where it has been shown that the Doppler 
shift of the absorption of weak 
lines provides a good measurement of the last scattering 
radius.  For non-spherical geometries the situation is even more 
complicated since the radial direction is not orthogonal to the shells of 
constant density.  For any definition, $R_{ph}$ 
becomes a function of the inclination angle $i$ of the line of sight
with respect to the direction 
along which the optical depth is measured. The latter is aligned with the 
density gradient, the direction of the highest escape probability for 
photons, and is taken to be the radial direction in the current work.  
Thus, $R_{ph}$ does not have an exact physical meaning but may be 
used for convenience to provide some guidance to imagine the physical 
situation.  Here, $R_{ph}$ is defined as the radial distance in the 
equatorial direction and we set $R_{ph}=R_{sc}$ since we are dealing with 
the interpretation of normalized spectra and $R_{sc}$ can be measured 
approximately by the Doppler shift of weak lines.
 
In conclusion, we use the following assumptions: a) ellipsoidal 
(oblate) geometry 
(Eq. 1); b) power law density profiles; c) velocity $v \propto r $; d) 
LTE-occupation numbers except for He; e) pure scattering approximation for the 
continuum and scattering lines with a thermalization fraction of $10^{-2} 
$; f) depolarization in lines and g) T(r) given by the gray extended 
atmosphere for the optically thick layers and T being constant in the outer,
optically thin layers.  
The free parameters are: a) photospheric radius $R_{ph}$, b) 
effective temperature $T_{eff}$, c) density exponent n, d) axis ratio E, e) 
enhancement factor for helium opacities.  
 
In the actual calculations, the angular space of the emitted photons is 
discretized by 20 bins in the inclination angle $i$ defined to be 
zero on the symmetry axis.  The spectral resolution is $\Delta \lambda / 
\lambda\approx 150 $.  The calculations are halted when the residual error 
in the percentage polarization P becomes less than 0.1 percent
(i.e. $P = 1.6 \pm 0.1\%$).
 
\section {The Calculations}
 
The large parameter space makes impractical a presentation of a grid of 
models covering SN~II in general.  Instead we have studied models 
applicable to SN~1993J at about 23 days past explosion.  We chose oblate 
ellipsoids with an axis ratio E of 0.6 and consider power law density 
profiles with exponents between --3 and --7 because those are expected
for the envelope of SN~1993J independent of the progenitor mass
or model (H\"oflich \etal\ 1993, Shigeyama \etal\ 1994, Woosley \etal\
1994).
Effective temperatures and the photospheric velocities range between 
4500 to 5500 K and 4500 to 5500 km s$^{-1}$, respectively.  The photospheric 
radius $R_{ph}$ follows directly from $v_{ph} $ and the time of 
observation.  In practice, we have adjusted the photospheric density 
$\rho_{ph}$ to produce $R_{ph}$.  We assume an equal amount of hydrogen 
and helium, a solar mixture of heavier elements, and for simplicity
we ignore composition gradients.  
The latter assumption
will be discussed in the conclusions.
Table 1 gives the 
parameters and some related quantities for those models that we discuss in 
detail.

Deviations from sphericity will result in a directional dependence of the
computed luminosity, i.e. $L=L(i)$. 
Consequently, the total luminosity cannot be 
deduced from observations without taking the asphericity into account.
This may have strong implications for the interpretation of 
light curves and the use of SN~II as distance indicators (H\"oflich 1991, 
1995).  This effect is illustrated in Fig. 1 for the current models.  
The small differences in $L(i)$
between the models with different density profiles
can be understood in terms of 
multiple-scattering of the continuum which decreases with increasing 
density gradients.  The variation of luminosity
with inclination will be a strong function of the ellipticity, E.
Although these models do not have a fully self-consistent temperature
structure, we believe this effect of inclination on luminosity
to be general.
 
\begin{table}[t]
\vspace*{5mm}
\caption{
Some characteristics of those models described in this paper in
detail. The quantities given in columns 2 to 7 are: The effective
temperature $T_{eff}$, the photospheric radius $R_{ph}$ and velocity
$v_{ph}$, the density at the photosphere $\rho_{ph}$, the index  n
of the power law density  $r^{-n}$, and the axis ratio E between
minor and
major axis of the ellipsoid. $R_{ph}$ is set to $10^{15} cm $ in all
models.}
\begin{tabular}{cccccc} 
\hline
  Model & $ T_{eff}$ & $v_{ph}  $ & $\rho_{ph}      $ &n & E  \\     
        &    [K]     & $[km/sec]$ & $[g/cm^3]$ &  &    \\     
\hline
H3cHe    &  4800 &  6600. & 6.55E-13 & 3  & 0.6 \\ 
H5cHe    &  4800 &  6600. & 2.25E-12 & 5  & 0.6 \\ 
H7cHe    &  4800 &  6600. & 4.35E-12 & 7  & 0.6 \\ 
\hline
H3sphere &  4800 &  5280. & 8.20  & 3  & 1.0 \cr
\hline
\end{tabular} 
\end{table}
 
\begin{figure}
\psfig{figure=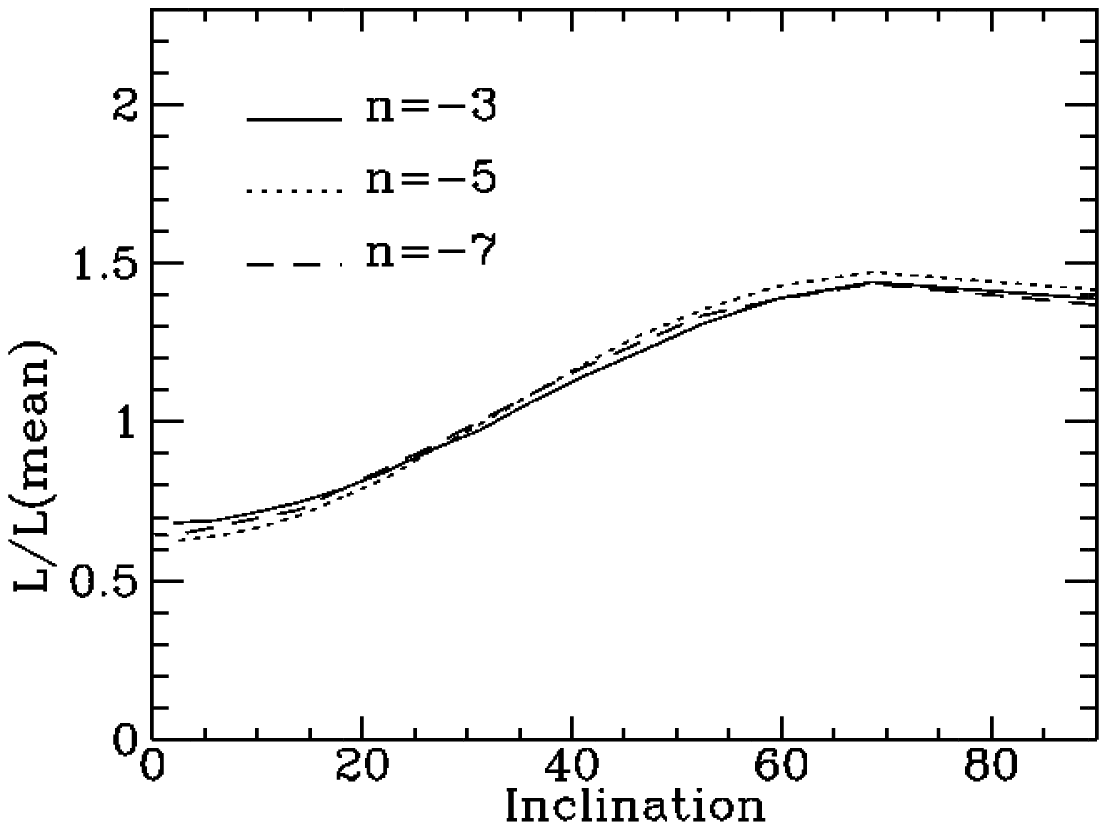,width=11.6cm,rwidth=4.5cm,angle=270}
\begin{small}
Fig. 1 --   Frequency mean of the luminosity as a function
 of the inclination angle for the models given in Table 1.
\end{small}
\end{figure}

\begin{figure}
\psfig{figure=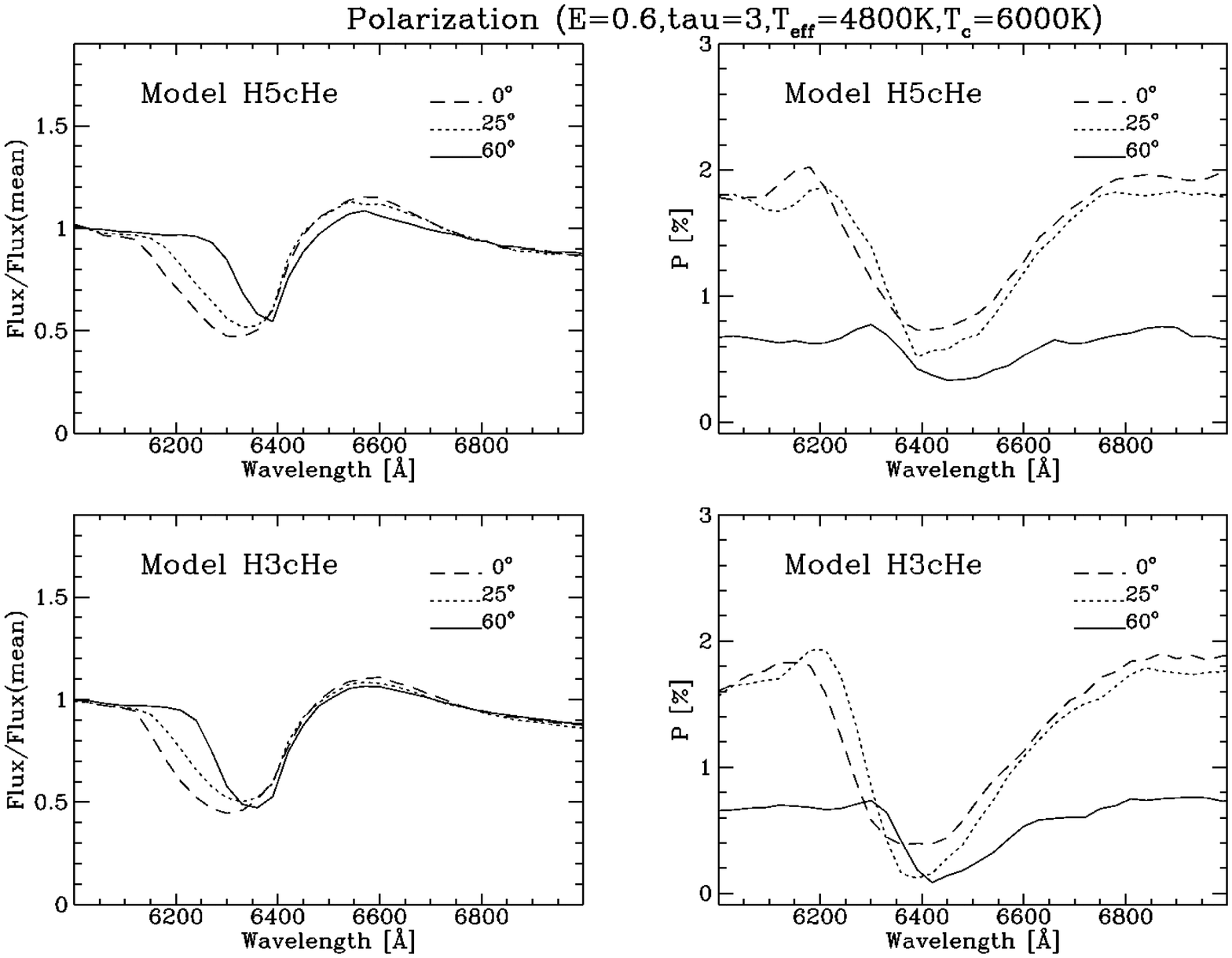,width=6.4cm,rwidth=9.5cm,angle=270}
\begin{small}
Fig. 2 --   Flux and  polarization spectra for different inclinations 
 for models  with n=5 (upper graphs) and n=3 (lower graphs). 
\end{small}
\end{figure}

The general properties of our model spectra can be discussed in terms 
of representative models.  We first consider models
with $T_{eff}=4800K$, $v_{ph}=5500\ km\ s^{-1}$, 
$R_{ph}=10^{15}cm$ and power law density gradients with n of 3 (H3cHe) and 5 
(H5cHe).  All flux spectra presented in the figures have been
normalized to the flux at 7000 \AA\ for convenience of presentation.

Before discussing the overall spectra, the region of H$\alpha$ is studied 
because the opacities are dominated by Thomson scattering and only a single 
line (assuming He I $\lambda$ 6678 to be negligible at this epoch).
In Fig.  2, the emitted flux and polarization spectra 
are given for inclination angles of 
90\deg, 65\deg\ and 30\deg, that is from looking directly along the
equator to nearly down the symmetry axis.

\begin{figure}
\psfig{figure=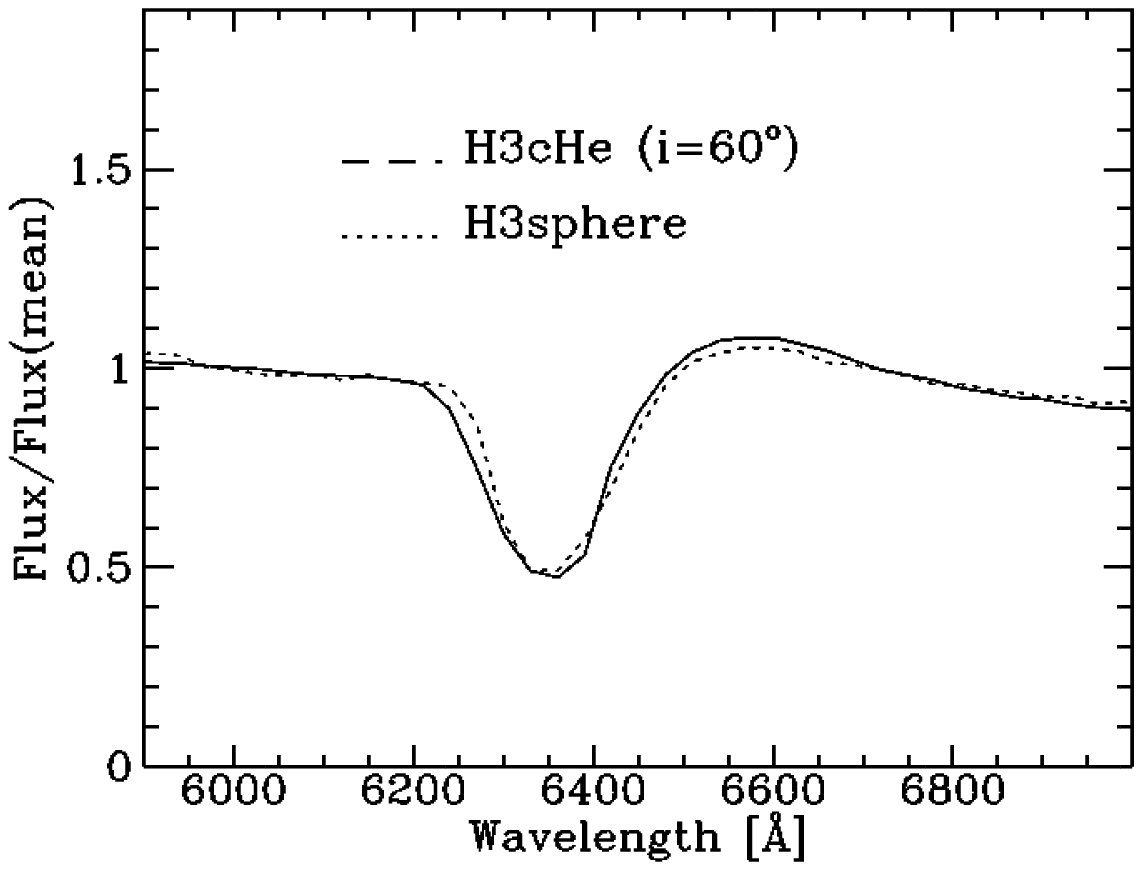,width=12.4cm,rwidth=4.5cm,angle=270}
\begin{small}
Fig. 3 -- Flux of H3cHe seen at an inclination of $30^o$ 
 in comparison to  a  spherical model 
with a $R_{ph}$ and $v_{ph}$ being reduced by 20 \% .
\end{small}
\end{figure}

\begin{figure}
\psfig{figure=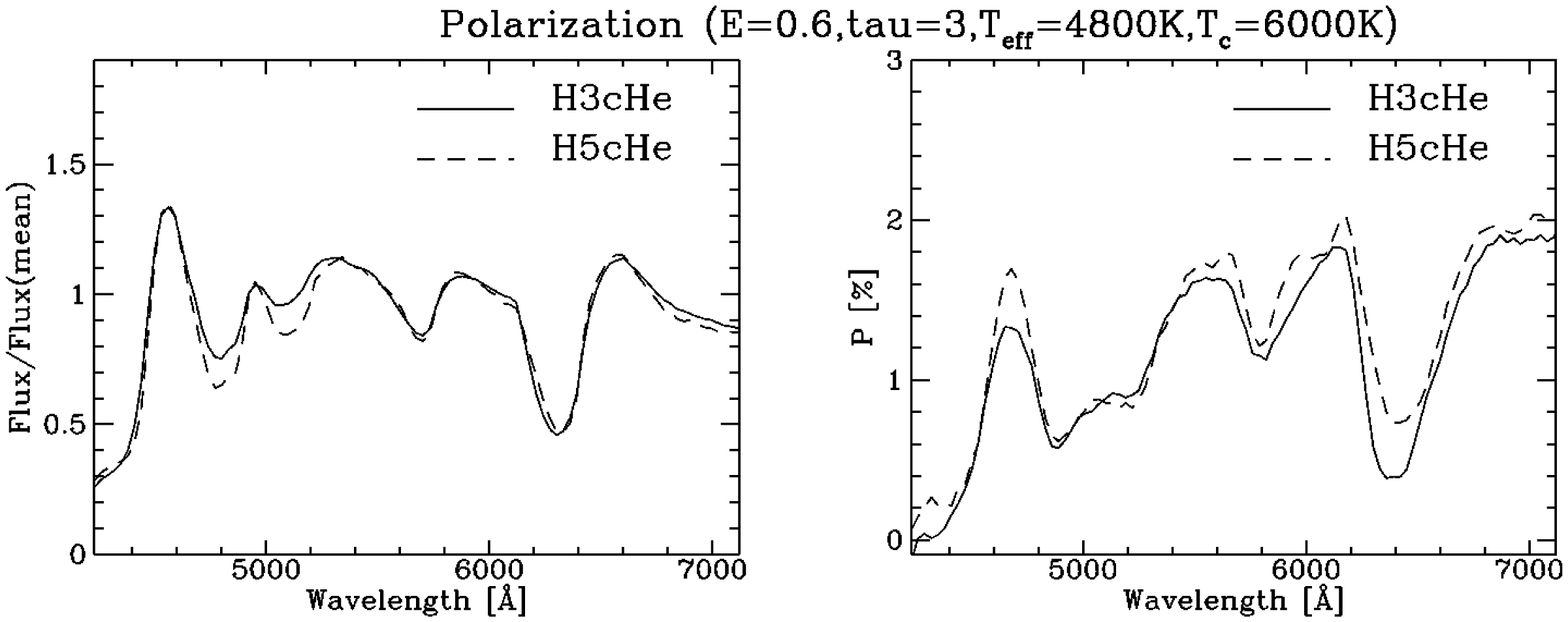,width=12.4cm,rwidth=9.5cm,angle=270}
\begin{small}
Fig. 4 --  Flux and polarization for H3cHe and H5cHe seen at $i=90^o$. 
\end{small}
\end{figure}

The line fluxes can be generally understood by analogy with spherical 
envelopes.  The density gradient determines the extension of the line forming 
region.  For homologous expansion, position translates into frequency or 
wavelength space.  Both the absorption and emission components become 
smaller with larger density gradients.  For optically thick lines, the 
absorption minima occur at expansion velocities in excess of the 
photospheric velocity $v_{ph}$, because they are formed well above the 
photosphere.  Thus, lines of different optical depth allow for a test of 
density gradients.  For a more detailed discussion of these and 
related questions see H\"oflich (1990,1991) and Duschinger \etal\ (1994).

The main effect of asphericity is a change of the Doppler shifts in lines
with the inclination i (Fig. 2).  The absorption component becomes sharper
and less blue shifted if observed from lower inclinations (i.e.  more
pole-on).  
For both H3cHe and H5cHe, the ratios between the Doppler shifts 
at $i=90\deg$ to that at $i=30\deg$ 
for both the absorption minima and blue edges 
are in the range 0.75--0.8.  This is consistent with the 
corresponding ratios of the radial distance of the photosphere $R_{ph}(i)$. 

We can construct a corresponding spherical model if we set
the photospheric radius for the spherical model equal to that 
$R_{ph}=R_{ph} (i)$ (Fig.  3).  
Even better agreement can be achieved if we would allow
for an artificial adjustment of the intrinsic line width by introducing a
``micro-turbulence" (H\"oflich 1988). We 
conclude that although the emitted flux is very dependent on the
inclination, observations of the flux spectra alone are not sufficient to
distinguish spherical and non-spherical geometries.
                                                             
Now, consider the polarization $P_{\lambda}$ (Fig.  2).  In the illustrated 
examples, $F_{\lambda}$ changes by only 50 percent over the H$\alpha$ 
feature whereas $P_{\lambda}$ varies by a factor of 3 to 10 depending on 
the inclination.  In general, the continuum polarization increases if the 
object is observed more directly on the equator (i.e.  from large 
inclinations $i$).  The polarization does not increase as $sin^2(i)$ as 
would be expected for optically thin envelopes.  The reason is that the 
depolarizing effects of multiple-scattering cause information on the global 
geometry to be lost thus reducing the emergent continuum polarization 
(H\"oflich 1991).  Because of the polarizing effects of Thomson scattering, 
we cannot assume that P$\rightarrow $0 even in the strong H$\alpha$ line 
where multiple scattering may be important.  This becomes significant is 
evaluating the effect of the ISM as discussed in \S\ 5.

As for the flux, the 
polarization profiles become narrower with increasing density gradients.  
Depolarization of a photon requires only one absorption in a line whereas
several interactions are required  
to change the flux $F_{\lambda}$ and the local source functions
by thermalization.  Even thermalization 
has little effect on the absolute flux over a line if, as we
assume here, the temperature 
gradients in the atmosphere are typically small.  
In addition, the depolarizing effect of multiple scattering 
is more important for radially extended atmospheres
The result, as illustrated in Fig. 2, is that $P_{\lambda}$ is much 
more sensitive to the density profile than is $F_{\lambda}$.

The breadth of the minimum in the polarization spectrum
provides additional information on 
the density gradient that is not available from the flux alone. 
For steeper density gradients, the particle density is higher 
at a given Thomson scattering optical depth (see Table 1).  For highly
ionized gas this implies that the number of neutral particles
is also higher at a given optical depth.
The Thomson optical depth depends linearly on the density, but 
the number density of neutral hydrogen formed by recombination
is quadratic in the density.  The result is that the   
line absorption of HI will be stronger compared to Thomson
scattering in a steeper density gradient.  
This will result in a stronger depolarization of the lines.  If H 
is mostly ionized, the bottom of the profile 
in the polarization spectrum
becomes slightly broader 
compared to its total width with increasing density gradient.

Changing the density gradient alters the flux and polarization
spectra in different ways at different wavelengths. This
is illustrated in Fig. 4.  For
temperatures larger than 4500 to 5000 K, line blocking due to singly 
ionized elements of the iron group increases with increasing density gradient 
due to the larger particle densities as just described. 
In the range of $T_{eff}$ under consideration, 
this effect works mainly at and below the 
photosphere.  In regions governed by a large number of weak lines, the flux 
is thus mainly determined by layers at 
large optical depths.  On the other hand, 
polarization is produced in layers of small optical depth since 
multiple-scattering reduces $P_\lambda $ at large optical depth.  
The resulting effect can be seen in the
spectra of Fig. 4 between 4800 and 5500 \AA.  The flux $F_{\lambda} $
of H5cHe is reduced in comparison to H3cHe,
whereas $P_{\lambda}$ shows little effect.  
Altering the density gradient alters the fluxes and
polarization in a different manner near H$\alpha$ than 
in the 5000 \AA\ region.  These differences 
provide a sensitive diagnostic tool.

Line blocking depends sensitively on such factors as 
$T_{eff}$ and the chemistry. Thus the models presented in 
this section should not be taken too literally.  Nevertheless,
these models should qualitatively 
represent a typical SN~II photosphere close to the recombination 
phase.
 
\section {Spectropolarimetry of SN 1993J}

SN 1993J was discovered on March 28.9 on the rise to its
first maximum.  It went through a minimum about April 5 and then 
a second maximum in V on April 18. 
Spectropolarimetry was obtained on 
April 3.2 and 4.2 by Bjorkman and Nordsieck
(1993) at the University of Wisconsin Pine Bluff 0.9~m telescope. 
Weighted over the spectral range 3200--7600 \AA, the results were 
$0.73 \pm 0.13$ percent at a position angle of $179.5\deg$ and
$0.22 \pm 0.16$ percent at a position angle of $41\deg$, respectively.  
The spread in these data is probably due to effects caused by the 
nearly full moon.  Formally combining the two data sets gives 
$0.46 \pm 0.1$ percent at a position angle of $5 \pm 6$\deg. 
Measurements made on April 7.22 UT by Smith (1993) yield
a $V$-band polarization of $0.19 \pm 0.11$ percent
at a position angle of $178 \pm 17$\deg. 
Observations of foreground stars along the direction to SN 1993J showed 
less than 0.1 percent
polarization, indicating that the observed polarization in 
SN 1993J is not caused by interstellar polarization within our Galaxy.

Spectropolarimetry was obtained at McDonald Observatory on April 20 
(Trammell, Hines, and Wheeler 1993).  The total polarization in the 
continuum was measured to be $0.9 \pm\ 0.1$ percent at a position angle of 
33 $\pm\ 3\deg$, nearly independent of wavelength.  The percent 
polarization decreased at the location of H$\alpha$ and the position angle 
rotated by about 10--15\deg.  On April 26, Jannuzi
\etal\ (1993) 
measured a polarization of 1.04 percent at PA = 29\deg\ with similar 
structure at H$\alpha$.  These results are consistent with those of 
Trammell et al. (1993) within the errors.  The polarization structure 
across H$\alpha$ alone proves that SN~1993J was intrinsically polarized, 
and that the lines and continuum had different polarizations regardless of 
any correction for interstellar polarization.

At H$\alpha$ the polarized flux consists of the underlying continuum of the 
supernova and the flux in the emission line, both affected by any 
polarization induced by the intervening interstellar medium (ISM).  After 
vector subtraction of the continuum polarization, Trammell \etal\ found the 
polarization of H$\alpha$ emission alone to be $1.1 \pm 0.1$ percent at a 
position angle of $150 \pm 4$\deg, similar in degree to the observed 
continuum, but with a significantly different position angle.  The 
orientation of this derived H$\alpha$ component is compatible with the 
position angle of the corresponding spiral arm on which SN 1993J is 
superimposed, suggesting that this component could be produced entirely by 
dust aligned by the magnetic fields that thread this spiral arm.  Radio 
Faraday rotation measurements support this interpretation since they imply 
that the local magnetic field closely follows the spiral arm containing 
SN~1993J at a position angle $\approx 150 $\deg\ 
(Kraus, Beck, and Hummel, 1989).
As mentioned above, the local Galactic contribution to
this polarization is expected to be minimal.
Trammell \etal\ assumed that H$\alpha$ was intrinsically 
unpolarized and that its derived polarization (after vectorially 
subtracting the continuum) was identical to that induced by the ISM in the 
spiral arm of M81.  They then isolated the intrinsic polarization of the 
supernova by correcting the total polarized flux for the interstellar 
contribution by means of a Serkowski law as parameterized by Wilking et al.  
(1982).  The result for the average continuum polarization of the supernova 
is then 1.6 $\pm$ 0.1 percent at a position angle of 49 $\pm$ 3\deg, both 
independent of wavelength.  The continuum polarization intrinsic to the 
supernova was thus found to be significantly larger than the total observed 
continuum polarization (i.e.  1.6\% vs.  0.9\%).  The 
position angle after correction for the ISM component 
is essentially independent of wavelength.
(Trammell \etal\ Fig.  1). This is evidence that the ISM has been properly 
isolated and subtracted since it is consistent with the source of the 
polarization being wavelength independent electron or
dust scattering (H\"oflich 1991, Jeffery, 1991).

After correction for the interstellar polarization within M81, Trammell 
\etal\ isolated the scattered continuum radiation by forming the Stokes 
flux.  This is done by rotating all the polarization into a single, rotated 
Stokes parameter then multiplying by the total flux (e.g.  Miller \& 
Goodrich 1989).  The intrinsic Stokes flux from the supernova computed in 
this way shows no peak at the rest wavelength of H$\alpha$, by construction 
(i.e.  H$\alpha$ is assumed to be intrinsically unpolarized).  The Stokes 
flux spectrum shows a broad minimum to the blue of H$\alpha$ (Fig.  2 of 
Trammell \etal).  This feature has a minimum at $\sim$ 6330 \AA\ 
corresponding to a velocity of 10,500 km s$^{-1}$ relative to the central 
wavelength of the emission line peak, and a FWZI of $\sim$ 395 \AA\ 
corresponding to a velocity range of 18,700 km s$^{-1}$.

Trammell \etal\ presented the percent polarization as a function
of wavelength for the total contribution, but did not present
the same quantity, the polarization spectrum as defined here,
corrected for the ISM.  These data are presented here (Figs. 6--8). The
polarization spectrum shows a broad, somewhat asymmetric minimum at about
6500 \AA.  

Spectropolarimetry data obtained on April 26 at Steward Observatory 
(Schmidt, 1995) and on April 30 by Tran \& Filippenko 
(1993) at Lick Observatory, both with excellent signal to noise ratio, 
confirm the basic interpretation of the spectra given by Trammell 
\etal.~~The Steward data, and additional Lick observations (Tran \& Filippenko 
1993b; Miller 1993), show a rotation of the position angle across H$\alpha$ 
that persists even after the ISP component derived by Trammell etal.  has 
been subtracted.  It is difficult to know whether such an effect is 
revealed by the superior signal to noise ratio or whether 
it is due to real changes in the polarized flux as the supernova evolves 
(both data sets were obtained during a period of rapid spectral evolution).  
This rotation does suggest a small polarization at H$\alpha$ 
intrinsic to the supernova.  This is possible because, as pointed out in 
\S\ 3, there can be some residual polarization induced by electron 
scattering in the line forming region. The different position angle 
compared with the continuum would then suggest slightly different 
scattering geometries for the emission lines and the continuum.

Note that if the ISM component has been correctly isolated by Trammell 
\etal\ as 1.1 percent at a 
position angle of 150\deg, then the early measurements 
of Bjorkman and Nordsieck (1993), and especially the ``null" result 
reported for April 7 by Smith (1993), $0.19 \pm 0.11$ percent
at a position angle 
of $178 \pm 17$\deg, require special attention\footnote{It should be noted 
that the numbers from Smith (1993) have been corrected for statistical bias 
via the prescription of Wardle \& Kronberg (1974), and that the reported 
numbers are consistent with zero observed polarization (e.g.  Simmons \& 
Stewart (1985)).}  The ISM value cannot change with time.  If there
is a finite contribution from the ISM then the supernova must contribute
in some way to produce a net null result.

Formal vector 
subtraction of the ISM polarization deduced by Trammell
\etal\ from the total polarization of the combined data of Bjorkman and 
Nordsieck and the total measured in $V$ by Smith yields a polarization for 
the supernova on April 3--4 of $1.0 \pm 0.1$
percent at a position angle of $48 \pm 6$\deg\ 
and on April 7 of $1.0 \pm 0.2$ percent at position 
angle 55 $\pm$ 6\deg.  This implies that the supernova 
already had an intrinsically polarized component on April 3--4
and on April 7, just before and just after the 
first minimum, respectively.  The data also suggest that the
polarization did not change appreciably in either amplitude or
orientation at that interesting early phase.

The close similarity of the position angles of the polarized
flux intrinsic to the supernova for these two
early data sets compared to the corresponding value derived on
April 20 by Trammell \etal\ of 49 $\pm$ 3\deg is remarkable.
Note that this angle, $\sim$ 50\deg, is 
significantly different from the position angle of any other component,
that is, the ISM.  This  
strongly suggests that the intrinsic polarization from the supernova was 
roughly constant or perhaps slightly 
increasing from $\sim$ percent 1 on April 
3--7 to $\sim$ 1.6\% on April 20 at essentially constant position angle, 
$\sim$50\deg.\footnote{We strongly encourage polarization observations of 
other stars within the spiral arms of M81 to independently estimate the 
interstellar component within the spiral arm.}  
Recently published photometric polarimetry by Doroshenko,
Efimov, and Shakhovskoi (1995) confirms this early trend
and reveals a mild wavelength-dependent decline in
polarization after the second maximum.

The spectropolarimetric data is thus a critical diagnostic for the models.  
The data of Jannuzi \etal\ (1993), Tran and Filippenko (1993), and Miller 
(1993) must be closely examined to check the degree of polarization and the 
position angle across H$\alpha$.

\section {Comparison with Observations}
  
The models can be compared with the observations of SN~1993J to show the 
potential of a simultaneous analysis of flux and polarization spectra and 
to show that the models may have some resemblance to reality.  
We seek qualitative agreement and 
restrict our considerations to quantities that will be little affected by 
NLTE.  Our comparison is based on the observational data of Trammell \etal\ 
(1993).   
We have assumed an 
interstellar reddening $E_{B-V} $ of $0.2^m$
substantially concentrated in M81
(Wheeler \etal\ 1993, Clocchiatti, \etal\ 1995). 

 \begin{figure}
 \psfig{figure=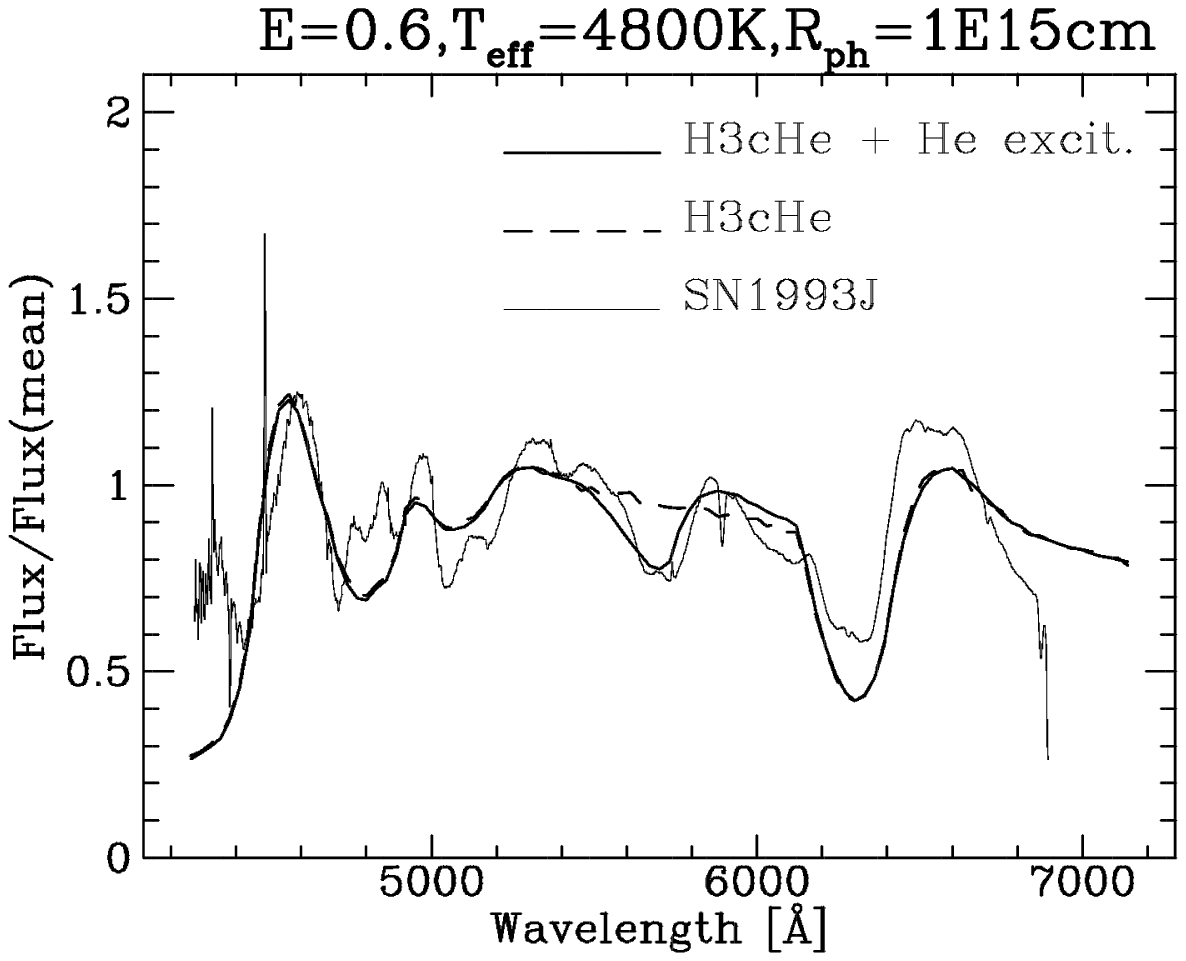,width=8.4cm,rwidth=9.5cm,clip=,angle=270}
 \begin{small}
 Fig. 5 --  Flux of H3cHe with and without and enhanced He I ($i=90^o$) in
 comparison to   observations of SN 1993J at April 21st, 1993.
 \vfill\eject
 \end{small}
 \end{figure}

The effective temperature is a free parameter in our models that 
determines the overall slope of the flux.  Best agreement with the 
observations is achieved for $T_{eff} = 4800 K$.  Models with $T_{eff}$ of 
4500 and 5500 K do not fit the overall flux distribution.  

Although the H$\alpha$ line may be subject to departures from
LTE, we treat it here in LTE.
The flux spectra show a flat-topped H$\alpha$ profile that
may be related to the imminent onset of the He I $\lambda$6678
line that becomes noticeble less than a week after the epoch
we consider or to the shell-like nature of the H envelope.
We do not attempt to reproduce this detail.  The absorption
component of H$\alpha$ shows evidence of a double minimum.
This may suggest some contamination from Si II $\lambda$6355
(Wheeler \etal\ 1994).  This feature may also be out of LTE
and we also do not attempt to reproduce the details of the 
absorption profile. 

The flux spectra also show a broad, perhaps double minimum
at about 5700 \AA.  This feature can be approximately reproduced
in the current models with a line of He I $\lambda$ 5876, but
only with a very large enhancement of the opacity of that
line, by a factor of $10^9$.
The He is expected to be far out of LTE due to the lack of
thermal photons and the presence of $\gamma $-radiation, but the required
enhancement factor seems excessive. 
 In their model for
the April 20 spectrum, Swartz \etal\ find a strong gradient
in the departure coefficients.  Typical values for the first
and second excited levels of He I (with higher levels having
somewhat smaller values are: $1.0\times10^{12}$ at the base
of the helium layer ($M_r = 0.7\ \m,\ v = 3300\ km\ s^{-1}$);
$1.0\times10^{8}$ in the middle
of the helium layer $(M_r = 1.5\ \m,\ v = 5800\ km\ s^{-1}$);
50 near the top  
of the helium layer $(M_r = 2.0\ \m,\ v = 7500\ km\ s^{-1}$);
and unity in the middle of the H/He envelope layer
$(M_r = 2.4\ \m,\ v = 7500\ km\ s^{-1}$). 
Although small variations in the electron temperature 
can change the LTE population numbers and hence the
normalized departure coefficients significantly, 
it thus seems unlikely that large departure coefficients 
and hence large variation from LTE opacities, of order $10^9$,
exist in the H envelope.

As shown by Swartz \etal\ (1993), a 
consistent treatment of the $\gamma$-ray excitation in the rate equations
gives a good representation of the spectra of 29 April and 9 May 
when the He I lines are clearly observed.
  The spectrum makes an obvious transition
between April 20, when the current spectropolarimetric data
were obtained, and April 29, when the He I lines begin to become
apparent.  In particular, the feature at 5700 \AA\ develops
a much stronger and sharp minimum that is clearly indicative
of He I $\lambda$ 5876.  It may be that the feature at 5700 \AA\
in the April 20 spectrum that we analyze here did not contain
any substantial component of He I, but rather comprised Na D
with perhaps some additional effect of Fe II lines.  Nevertheless,
in the current calculations the He I line has been included
and its effects on the polarization can be addressed.  The
qualitative conclusions may also pertain if the line were
instead Na D, probably also subject to NLTE excitation.

The theoretical spectra of H3cHe are compared 
to the observations in Figs. 6a-c.  
By comparing the flux spectrum $F_{\lambda}$ (upper panels)
with the polarization spectrum $P_{\lambda}$ (lower panels),
one sees that the depolarizing effects of the H$\alpha$ line
are distributed over the whole P-Cygni line profile.  
In both the observations and theory, the
depolarization sets in steeply somewhat to the red
of the blue edge of the absorption, reaches a minimum   
slightly to the red of the absorption minimum
and then rises more slowly through the rest frame wavelength
P-Cygni peak and into the red wing of the line.
The polarization profile is ``out of phase" with respect to
the flux profile since depolarizing requires only a single
line scattering but several scatterings are required to
substantially alter the flux. 
At the H$\alpha$ minimum of the polarization spectrum
$P_{\lambda}\sim 1$ percent.  This value represents 
the continuum polarization since the process of correcting
the observed spectra for effects of the ISM
adopted by Trammell \etal\ involved the assumption 
that there is no polarization in the H$\alpha$ line
itself at the rest wavelength. 
 
If viewed from the equatorial plane ($i=90\deg$), the theoretical 
absorption components of the line fluxes for model 
H3cHe are significantly too broad, 
although at the right frequency shift.  In general, the polarization 
spectrum provides the right amplitude for the percentage polarization, 
$P_\lambda$.  The blue 
wing of H$\alpha$ extends too far by about 200 \AA.  
The depolarization over the ``He I" line at $\sim$5700 \AA\  
is about correct.  For this model, both 
$P_{\lambda}$ and $F_{\lambda}$ show the need for smaller inclination.

At $i=30\deg$ (Fig.  6c), the flux profiles tend to be too narrow and the 
Doppler shift of the absorption component becomes too small.  The 
differences in the flux spectrum may be corrected by choosing a larger 
outer radius (i.e.  a somewhat larger expansion velocity), 
but then the amplitude of $P_{\lambda}$, which is already
too small in Fig. 6c, would decrease even further due to 
enhanced multiple scattering.  

For oblate models, the depolarization 
in H$\alpha$ increases as the inclination decreases
because as the line of sight moves toward the symmetry axis
there is a steeper density gradient and hence an
increase of the depolarizing effects of multiple scattering
compared to Thomson scattering.  This effect will occur
at any density gradient, but its amplitude will depend
on the value of the density gradient.
In Fig. 6c, $P_{\lambda}$ at the absorption minimum of H$\alpha$
essentially vanishes, in strong disagreement with the observations.
This is because, in this model, the optical depth for line scattering
is so high compared to Thomson scattering that even the
residual continuum polarization from the more nearly 
symmetrical geometry is removed by depolarizing effects of 
line scattering.

In principle, the absolute polarization in the continuum 
could be increased with larger asphericity.  The fact that 
$P_{\lambda}\sim0$ 
at the absorption minimum of H$\alpha$  
is not likely to change with moderate increase of asphericity
and hence this problem can not be fixed in that way.
The excessively small polarization at H$\alpha$ might
be adjusted if the adopted polarization of the ISM were incorrect.
As discussed below, there is no evidence that the applied
correction is substantially in error and altering that
correction would not change the relative degree of depolarization
at H$\alpha $ compared to shorter wavelengths. In fact, the
depolarization at wavelengths below 5400 \AA\ is too weak compared to 
H$\alpha$ and this problem would remain if there were
a simple renormalization associated with a different ISM polarization
component.  

Thus, although some parameters could be adjusted to improve certain 
aspects of the calculation, we conclude from the model H3cHe
that small inclinations are 
not consistent with
SN 1993J.

A medium inclination $i=65\deg$ alleviates some problems with respect to the 
flux spectra, but suffers both the problems with excessive
depolarization in H$\alpha$ that plague $i=30\deg$ 
and the 
overly broad absorption components in 
$F_{\lambda}$ that occur for larger $i$.  
Overall, models with excessively low density gradients,
n $\lta$ 3, are not acceptable.
\begin{figure}
\psfig{figure=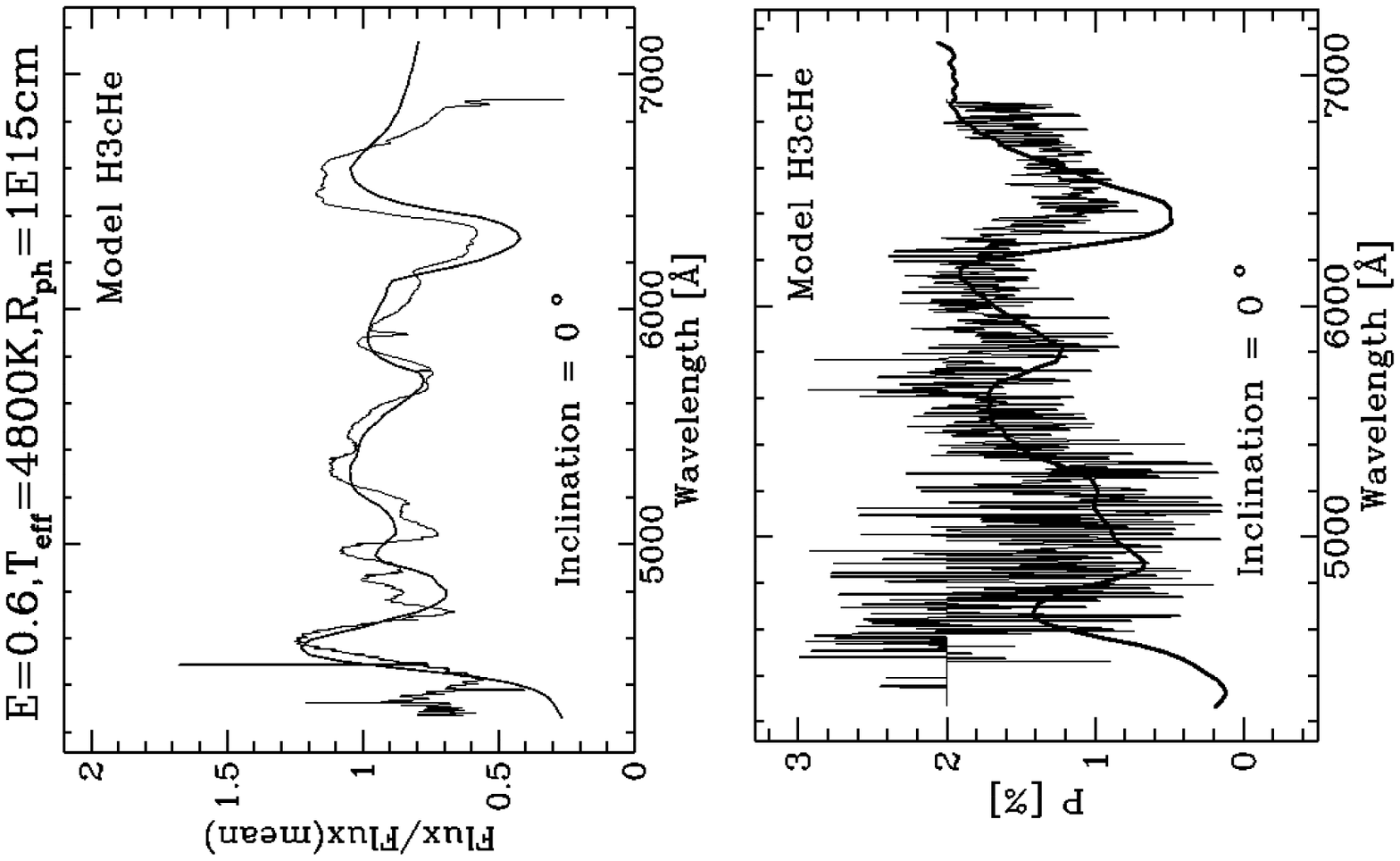,width=12.6cm,rwidth=9.5cm,angle=270}
\begin{small}
Fig. 6a --  Flux of H3cHe seen at  an inclination of $90^o$ in comparison
with the spectrum  of SN 1993J observed at April 21, 1993.
\end{small}
\end{figure}

\begin{figure}
\psfig{figure=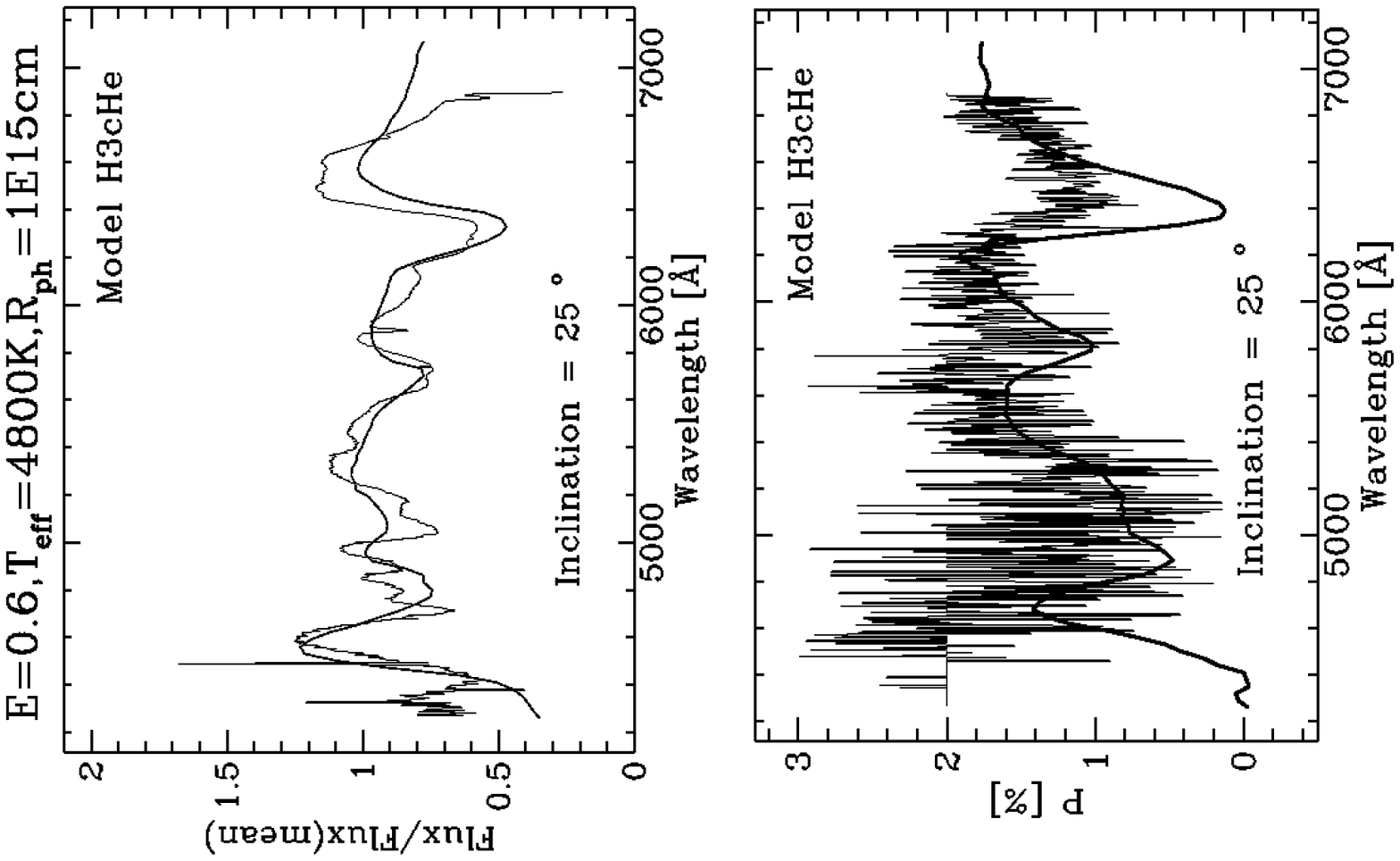,width=12.6cm,rwidth=9.5cm,angle=270}
\begin{small}
Fig. 6b --  Same as Fig. 6a but  $i=65^o$.
\end{small}
\end{figure}

\begin{figure}
\psfig{figure=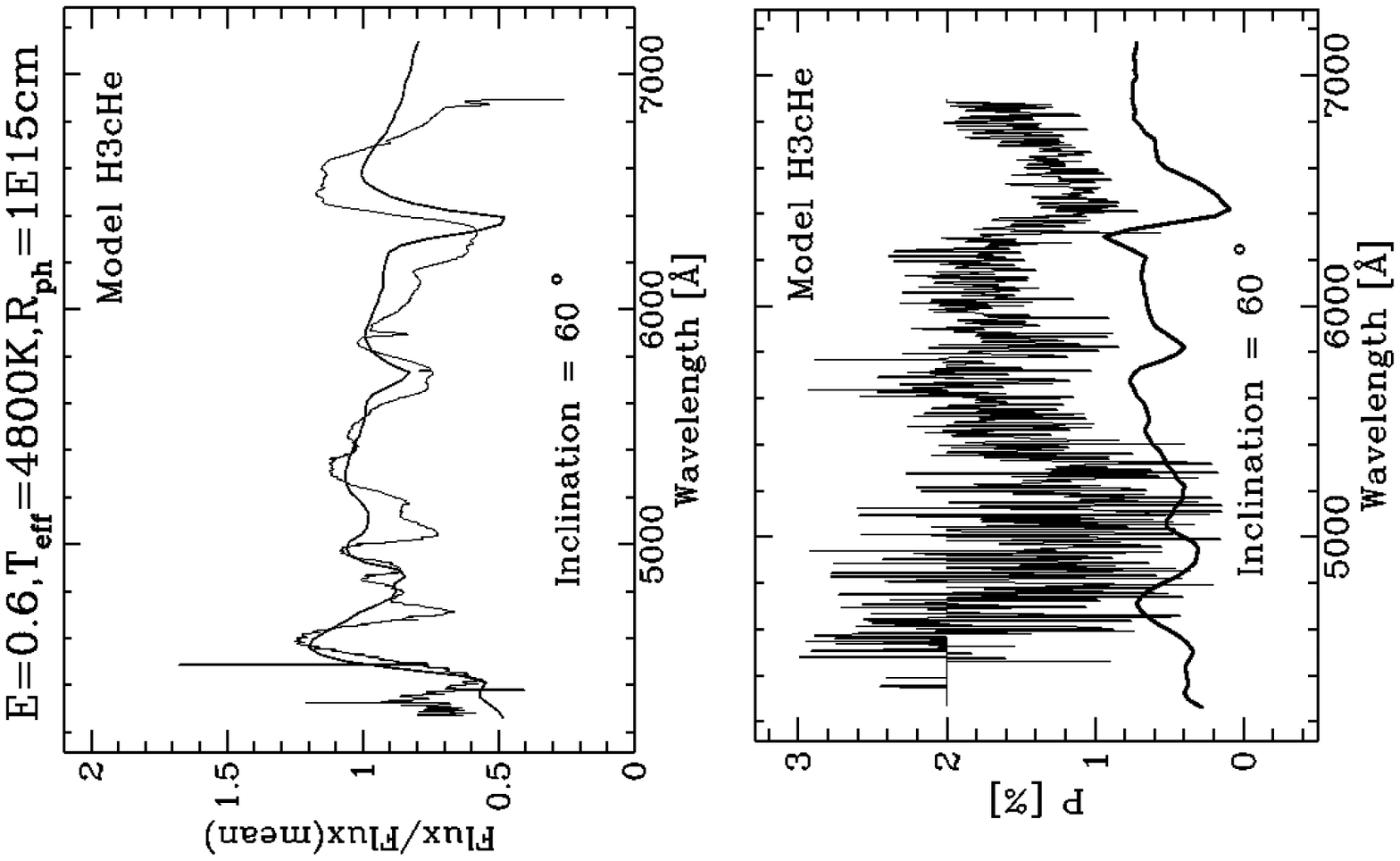,width=12.6cm,rwidth=9.5cm,angle=270}
\begin{small}
Fig. 6c --  Same as Fig. 6a but $i=30^o$.
\end{small}
\end{figure}

\begin{figure}
\psfig{figure=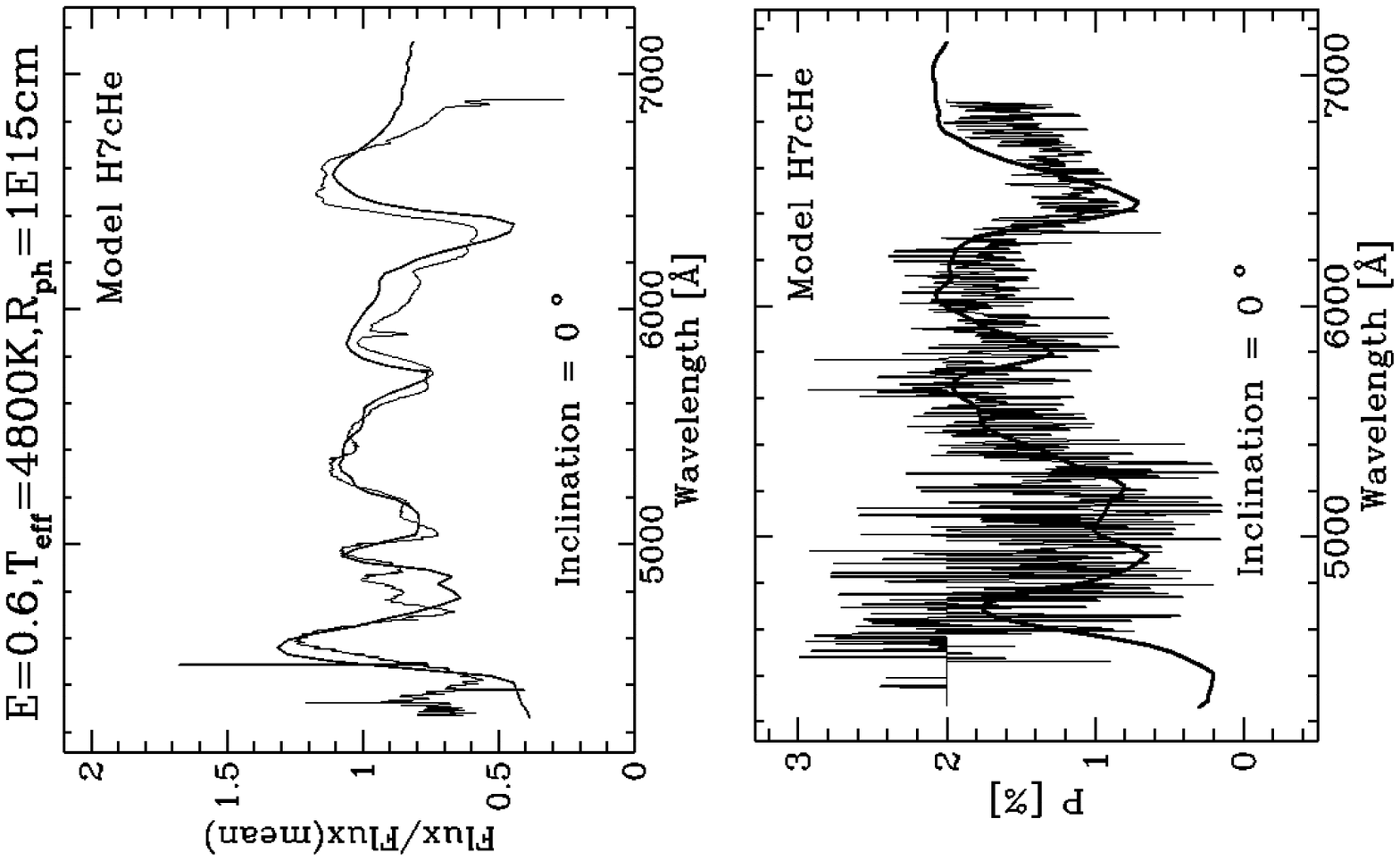,width=12.6cm,rwidth=9.5cm,angle=270}
\begin{small}
Fig. 7  -- Flux  H7cHe seen at an inclination of $90^o$  in comparison
with the  spectrum of SN 1993J at April 21, 1993.
\end{small}
\end{figure}

For the flux spectra, better
agreement with the observations can 
be achieved for power law densities with n = 5 to 7 (Figs. 7 and 8abc).  
Further adjustment of parameters might result in closer agreement. We forego 
this exercise because the limits of the models would 
prevent gaining much additional information.

\begin{figure}
\psfig{figure=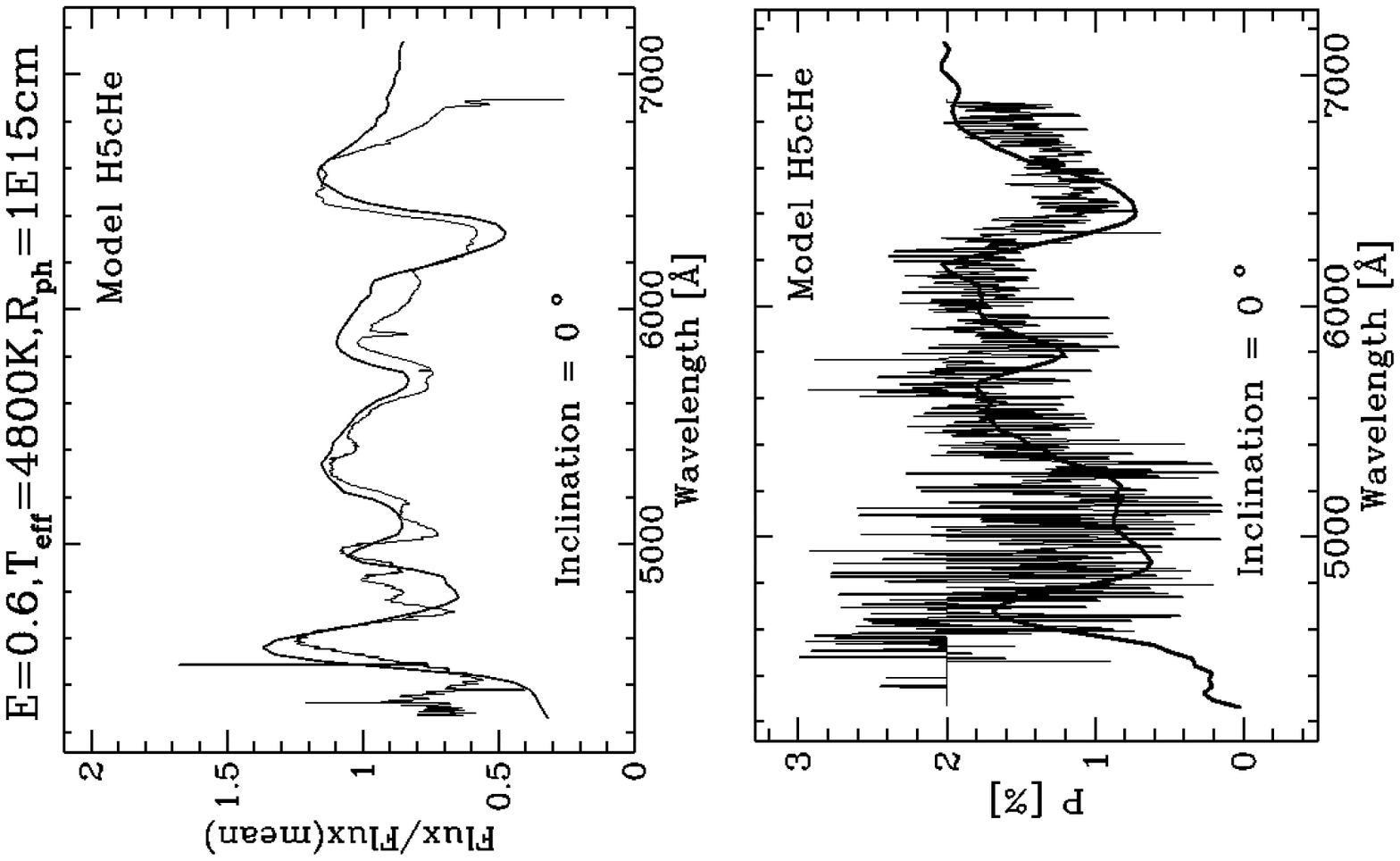,width=12.6cm,rwidth=9.5cm,angle=270}
\begin{small}
Fig. 8a -- Flux  H5cHe seen at an inclination of $90^o$  in comparison
with the  spectrum of SN 1993J at April 21, 1993.
\end{small}
\end{figure}

\begin{figure}
\psfig{figure=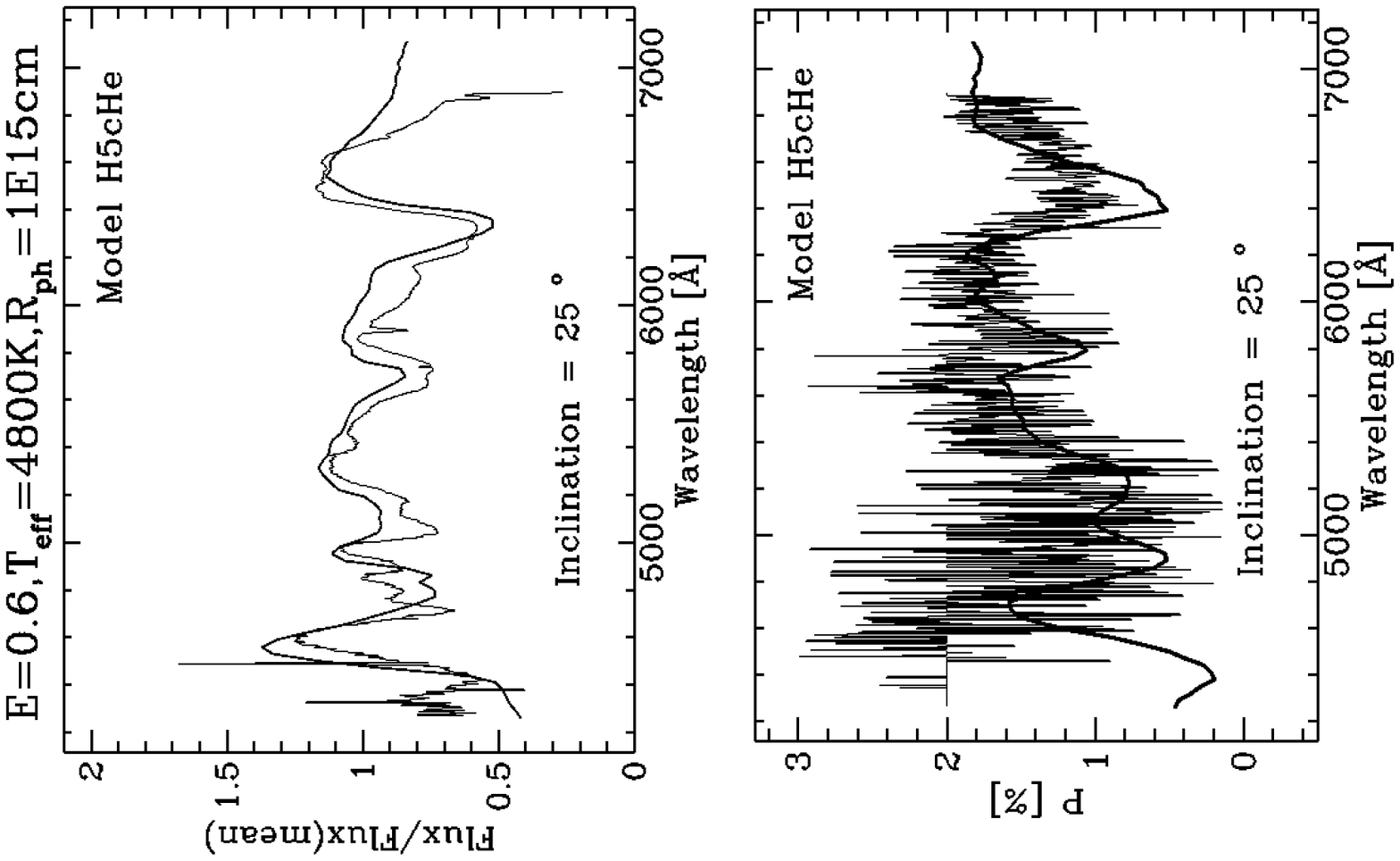,width=12.6cm,rwidth=9.5cm,angle=270}
\begin{small}
Fig. 8b --  Same as Fig. 8a but  $i=65^o$
\end{small}
\end{figure}

\begin{figure}
\psfig{figure=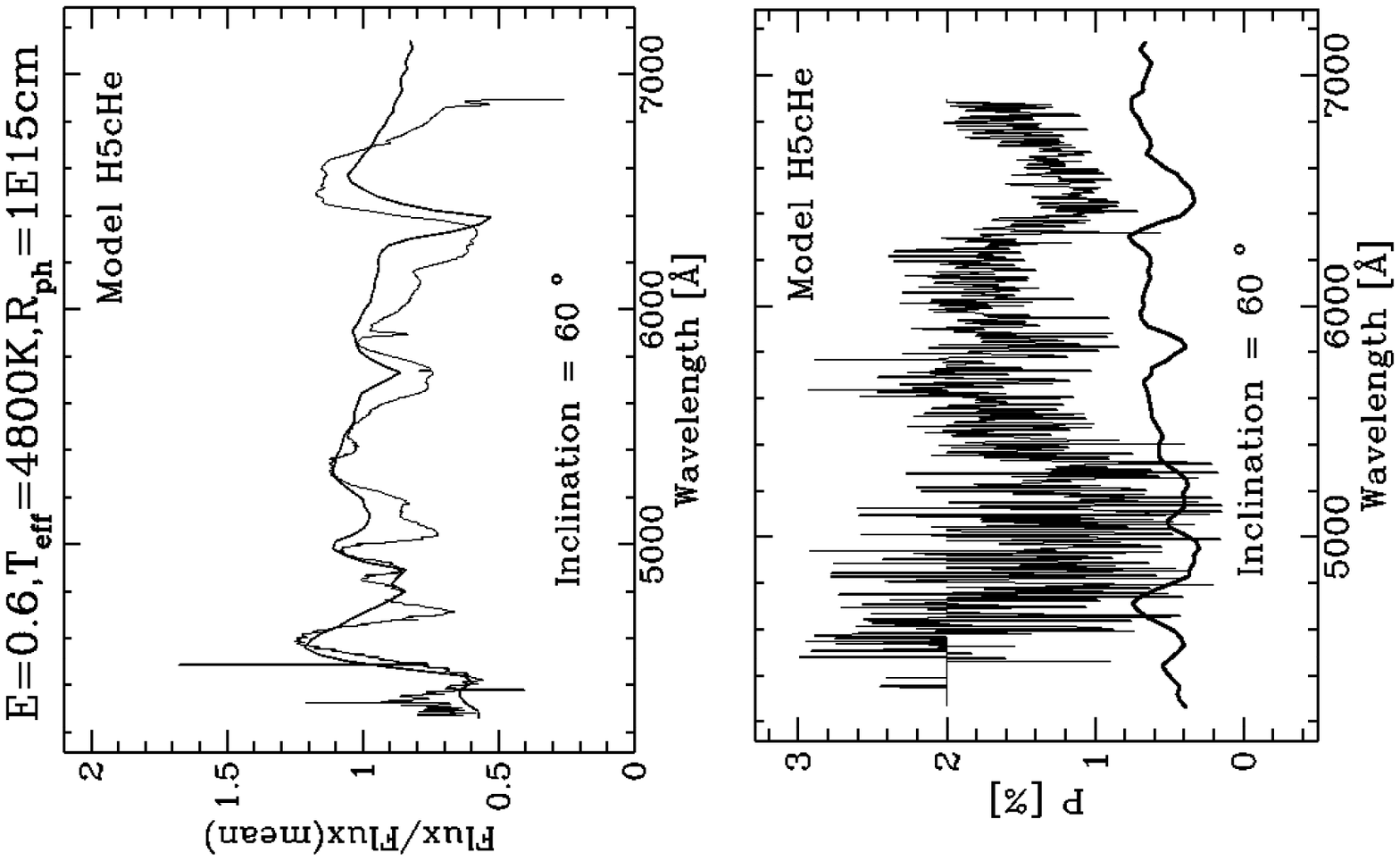,width=12.6cm,rwidth=9.5cm,angle=270}
\begin{small}
Fig. 8c --  Same as Fig. 8a but  $i=30^o$
\end{small}
\end{figure}

The polarization spectrum shows that H7cHe also has some problems (Fig. 7).  
In particular, H$\alpha$ is significantly too narrow by almost a factor of 
two.  For steep density gradients, the wavelength range with 
significant optical depth is very restricted. 
This problem becomes more severe for models with even 
steeper density gradients and corresponding smaller scale heights.
The polarization spectrum, in particular, puts an upper limit on
the density gradient in SN~1993J.  Models with $n\gta7$ 
do not give acceptable agreement with the observations.

A reasonable fit to both the flux and polarization spectra is provided by 
H5cHe at high inclinations.  Since the absolute degree of depolarization 
becomes too strong for low inclinations (see above), we conclude that the 
inclination 
should be larger than about 65\deg and is probably close to 
90\deg.  With this combination, both $F_\lambda$ and 
$P_\lambda$ for SN~1993J are reproduced reasonably well.
 
\section {Discussion and Conclusions}

A new version of our Monte-Carlo code is presented which allows the 
calculations of polarization and flux spectra for arbitrary 3D-geometries 
in rapidly expanding envelopes if the temperature, velocity and density 
structure are given.  Calculations using this code for 
scattering dominated, rapidly expanding photospheres have been performed to 
study the influence of deviations from sphericity on the spectra and 
polarization of 
SN 1993J.
\subsection {Implications for Distance Determinations}

We have shown that deviations from sphericity as high as a factor of 2 
are difficult to 
detect by analysis of spectral fluxes alone.  Although the flux 
depends sensitively on the direction from which an envelope is observed, it 
is possible to find a set of parameters for a corresponding spherical model 
that reproduces the observations.  Note that a similar result has been 
found for static, isothermal envelopes by Schmid-Burgk (1982) for the 
continuum problem.  

Since observed spectra of an aspherical supernova 
can be reproduced by spherical
models with the adjustment of free parameters such as the stellar radius or
by introducing a small microturbulence, we can not draw accurate 
conclusions about the geometry of a given supernova
based solely upon the flux density 
spectrum. Even in the case of 
very well observed supernovae such as SN 1987A, asphericities of the order 
of 10 to 20 percent may remain undetected if no polarization measurements are 
available.  This conclusion has implications not just for our understanding of 
supernova explosions and the progenitor evolution, but also 
for the use of SN~II as distance indicators.  Asphericity intrinsically 
demands a direction dependence of the deduced intrinsic luminosity.  
The application of the Expanding Photosphere Method to Type~II
supernovae in particular must treated with care.

\subsection {Effect of Density Gradient and Inclination}   

The sensitivity of polarization spectra to the underlying density structure 
and the inclination angle $i$ has been demonstrated.  The polarization 
provides a good test of the optical depth in a line, since photons are 
depolarized even by moderate $\tau$.  On the other hand,
the intrinsic polarization 
does not necessarily vanish over strong
 lines such as 
H$\alpha$ because of Thomson scattering within the line forming region.  
Depolarization by many lines in a broad wavelength range is as effective
as a single strong line, but regions of moderate optical depth in
the individual weak lines are more geometrically compact. 
This, in turn, lowers the effective
contribution of Thomson scattering and the residual polarization.
The ratio between depolarization caused by single, strong lines to that by 
multiple lines of moderate opacity thus provides additional information, e.g.
about the density structure.

The observations of SN 1993J have been 
reanalyzed  using the polarization and flux 
spectra of Trammell \etal\ (1993). 
For the oblate models studied here,
we find that the best agreement with observations
occurs for power law index $n$ of the density 
gradient ($\rho\propto r^{-n}$) at the line forming region  
in the range 3 $\lta n \lta 7$ with a best value of
about n = 5. 
The flux spectra also suggest that the density gradient cannot
be too steep, but the polarization gives more stringent
limits. 

Although
we have not varied the ellipsoid axis ratio E in this study, a value 
of 0.6 produces results in reasonable quantitative
agreement with previous investigations (H\"oflich 1995).  
If oblate ellipsoidal geometry holds for the envelope structure of SN 
1993J, we can infer that SN 1993J was seen ``equator on." 
Oblate models give more polarization
when viewed on the equator for two reasons that act in the
same direction.  The first reason is that the geometrical effect of the 
distortion from spherical symmetry is maximized along the equator. 
The second effect is that the depolarizing
effects of line scattering versus Thomson scattering are minimized 
for equatorial lines of sight
due to the shallower density gradient and lower photospheric density.

We have not explicitly studied prolate geometries here.  
For prolate models, the effect of geometry on polarization
is again maximized by
viewing along the equator.  The depolarizing effects of
line scattering are, however, also maximized because now the steepest 
density gradients and photospheric densities are along the
equatorial, not the polar direction.
This is why, in general, such 
prolate geometries
would require higher distortions (smaller values of the 
axis ratio E) to reproduce a given degree of polarization.  
Quantitative constraints on 
the density gradient and inclination angle of prolate
models would require specific numerical models which we have not
done here.  We believe that  
adopting prolate geometries would not
substantially alter our conclusion that density gradients
can be neither too steep nor too shallow, nor our conclusion
that SN~1993J was observed approximately normal to the symmetry axis.
 
\subsection {Determination of the Interstellar Polarization Component}
 
The problem of the interstellar component of the polarization can be 
re-examined in the context of our results.   The procedure
adopted by Trammell \etal\ to evaluate the ISM component 
involved the formal assumption that the polarization at H$\alpha$
consisted of supernova continuum, an added Serkowski-law component
imposed by the ISM in M81, and the contribution from H$\alpha$ itself.
The implicit assumption was that a polarized continuum emerged
from a deep layer, then passed through the H scattering layer,
and then passed through the ISM.  This allowed the three
components to be combined
by vector linear addition.  In reality the 
interaction of the continuum and lines
is more complex.  Trammell \etal\ took the intrinsic polarization
of the H$\alpha$ line to be zero and assumed that 
the intrinsic continuum polarization
from the supernova is continuous so that its
value at H$\alpha$ could be determined by interpolation from
the surrounding continuum.  This allowed them to subtract
the continuum contribution at H$\alpha$ and to assume that
the remaining polarization at H$\alpha$ gave the amplitude
and orientation of the ISM. 

The assumptions that the continuum can be interpolated and 
that only continuum electron scattering and the ISM contribute
at H$\alpha$ may not be correct [e.g. Schmidt 1995; 
Tran \& Filippenko (1993)].  
We see from the models, especially those
that are viewed nearly pole-on, that the strong depolarizing
effects of the line scattering can depolarize the continuum 
so that the continuum at line center cannot be evaluated
by interpolation.  In other circumstances, some continuum polarization
may be added at H$\alpha$ since scattering electrons are
present in the line scattering region.  The continuum
polarization is not necessarily determined in a layer
completely beneath the H layer.  

Although these caveats must be carefully considered and re-evaluated
if new data become available, the current models suggest that
the ISM has been correctly removed from the polarization spectral
data presented here (Figs. 6--8).  These models reproduce the 
line profiles in terms of the breadth of absorption components, the amplitude 
compared to the continuum, and the line shifts for both $F_{\lambda}$ and 
$P_\lambda$.  The differing amounts of depolarization between different 
wavelength regions can also be reproduced.  In addition to the flux 
distribution given by $T_{eff}$, we have adjusted $v_{ph}$, n and $i$.
The ellipsoid axis ratio E is given by the absolute size of 
$P_{\lambda}$ and $R_{ph}$ follows directly from $v_{ph}$ and the time of 
explosion.  Significantly larger distortion 
(smaller E) would require a smaller 
inclination angle $i$ which can be excluded from the polarization spectrum
over H$\alpha $.

We conclude that the ISM correction suggested by Trammell \etal\ (1993) is 
likely close to reality.
Determining the interstellar component from a 
measured polarization may be difficult in general 
without additional information such as the position angle
as a function of wavelength 
and, as for the case of SN 1993J, the magnetic field structure within the 
spiral arms of M81.  In principle, the technique 
outlined in this paper can be used to iteratively determine the 
interstellar component independently of other estimates.

\subsection {Constraints on Dynamic Models}
 
This analysis of the polarization allows us to re-examine the
dynamical models for SN 1993J in a new perspective.
One of the significant controversies about the nature of the 
explosion in SN~1993J is the question of when the photosphere
receded from the outer hydrogen layer into the inner helium mantle.
The various points of view are summarized in Wheeler \& Filippenko
(1995).  In models with small ejecta mass, $\sim$1.6--4.6 \m,
modest hydrogen envelopes, $\sim$0.1--1 \m,
and little outward mixing of radioactive nickel,
the photosphere recedes into the helium layer at the first
minimum of the light curve about seven days after the explosion
(Wheeler \etal\ 1993; Woosley \etal\ 1994).  The
models of Wheeler \etal\ with no nickel mixing
did a good job of matching the V
magnitude through the second peak, but found the photosphere to rapidly
recede through the relatively cold and neutral helium and into
the C/O mantle.  This resulted in too rapid an evolution to
the blue after the first minimum.  
Woosley \etal\ invoked a modest mixing of Ni outward
and the photosphere was held in the helium layer until near the
second maximum.  Models by Shigeyama \etal\ (1994) and 
H\"oflich \etal\ showed the photosphere lingering in the
hydrogen envelope until near the second peak.  Shigeyama \etal\
invoked an ejecta mass comparable to that of Wheeler \etal\
and Woosley \etal\ but assumed a considerable amount of Ni
mixing.  This may not be allowed since if the hydrogen is
too strongly exposed to the radioactive decay, it will be
over-ionized and the H lines too weak (Swartz \etal\ 1993).
H\"oflich \etal\ have explored models at the extremes of
rather low ($10 M_\odot$)  
and rather high ($30 M_\odot$) main sequence mass 
and found that the low mass     
models were strongly contradicted by both the observed 
spectra and light curves due to the lack of a massive inner C/O core.
The more massive models of H\"oflich \etal\ contained about ($\sim$ 11 
and 3 \m\ in the ejecta and hydrogen envelope, respectively,
and also invoked extensive outward mixing of Ni.  
The initial and final mass ejected by SN1993J is probably somewhat
lower (see below).

There are several reasons to believe that, despite other
possible quantitative problems (e.g. interstellar reddening),
the models of Shigeyama
\etal\ and H\"oflich \etal\ are qualitatively correct
in the statement that the photosphere did not recede into
the helium layer until near the second maximum.  One of these
is the color evolution.  The B-V color reddened during the 
first decline to minimum.  The color was relatively flat
during the subsequent rise to the second maximum at
which point it evolved rapidly redder again.  The color plateau
on the second rise is not easy to explain.  Wheeler \&
Filippenko suggest that it might represent a phase when
the photosphere is trapped in the helium layer (by radioactive
heating) and that the subsequent reddening is when the photosphere
finally recedes into the metal core.
This may not be consistent with the spectral evolution.  
The helium lines were not observed until just before the second
maximum (Swartz \etal\ 1993, Filippenko, Matheson, \& Ho 1993).  
It is possible, but probably not likely, that the photosphere had receded into
the helium during the rise, but not sufficiently that strong
helium lines were produced.  Such a situation would seem
to require too much ``fine-tuning."  The  
spectral models of Swartz \etal\ require the photosphere to
be in the hydrogen layer on April 20, when the current spectropolarimetry
was obtained, and then recede into the helium layer by April 29
when the He I lines were distinctly observed in the optical and IR.
Utrobin (1995) gets reasonable results for the emergence of the
helium lines with a model
of 2.4 \m total ejecta of which the outer 2 \m is 
He fully mixed with about 0.1 \m of H.  There is
no pure helium mantle, in contrast to the results of 
most evolutionary calculations. 
In these models, the nickel is confined to the inner layers
of heavy elements. Despite lingering problems with the 
color evolution on the second rise, Utrobin's models provide an interesting
alternative to high mass hydrogen-rich envelopes and combine
some of the advantages of both the high mass and low mass models
outlined above.

The polarimetry gives another perspective on this issue.  If the
analysis of the results 
from Bjorkman \& Nordsiek (1993) and from Smith (1993)
presented here (\S\ 4) are taken at face value then the polarization
did not change its orientation nor value at the time of the
first minimum.  This suggests that there was not a major change 
in the structure of the atmosphere at that time.  Rather than
a recession of the photosphere into the helium layer, this
may have just been a time of transition from adiabatic cooling
by shock heating to heating by the interior radioactive decay. 

The quantities $P_{\lambda}$ 
and $F_{\lambda}$ measure different physical effects (see above) and, 
consequently, different layers of the expanding envelope.  
Both the polarization and flux spectra presented here 
are consistent with the photosphere still being in the
hydrogen layer on April 20, three weeks after the explosion.  
The models are consistent with 
nearly the same expansion velocity 
for the H$\alpha$ and ``He~I" lines despite
rather different optical depths, $\sim$ 100 and 10 for H$\alpha$
and He I, respectively.  This evidence for homogeneity of composition
may need re-evaluation if the relevant feature is not He, but Na. 

Test calculations in which He has been substituted for H up to $ 2 
R_{ph} $ 
to test whether the photosphere could be in a helium mantle
in late April gave poor results.  The H profile becomes overly
boxy in both absorption and emission and in the polarization
profile.  The amount of polarization also drops because there
are fewer free electrons to scatter in the pure helium layers.
This could be overcome by increasing the distortion of the geometry,
but this would then demand a correspondingly larger polarization
in the earlier observations when the photosphere was definitely
in the H envelope. It seems unlikely that the distortion grows inward,
but perhaps this cannot be ruled out.  
The models in which the photosphere
remains in the hydrogen layer until after day 20
are qualitatively consistent
with the small increase in derived polarization since the
photosphere will recede to deeper layers where the density
gradients tend to be less and hence depolarization is decreased.
Qualitatively, the present results  
agree with those of Swartz \etal\ 1993 (see also
Baron \etal\ 1993, 1994) that on April 20 the photosphere 
is still formed in the H/He rich layers.

The models presented here put constraints on the gradient
of the density at the photosphere.  
The present results show that a gradient
steeper than n = 7 is not consistent with the current models and
observations and that n = 5 is preferred.
This puts constraints on the structure of the dynamical model
and perhaps on the interaction with the circumstellar medium.
The explosion of a star with an initial structure resembling
a polytrope of index 3 will give a post-shock structure
with a slope n $\sim$ 7.  On the other hand an initial structure
with a fairly massive (comparable to the core mass) extended
envelope will yield a post-explosion structure with the
envelope compacted into a shell with a very steep density gradient
on the outer face.  Models of SN~1987A gave density gradients
with n $\sim$ 3 near the base of the H envelope.  Moderate
mass models for SN~1993J (Shigeyama \etal\ 1994) give
n $\sim$ 20--25 in the very outermost layers, but closer
to n = 3 in the bulk of the H envelope and into the helium mantle.
Clocchiatti \etal\ 1995 conclude that the photosphere did
have a steep density gradient in the very early phases, but
the current observations pertain to several weeks after maximum.
There dynamical models are more consistent with a modest
density gradient as we have derived here.
Baron \etal\ (1993, 1994) consider exponential atmospheres
and suggest that the gradient must have an equivalent density index
of n = 12 -- 23 at this phase.   This does not seem to be consistent
with the dynamical models nor the results derived here.

\subsection {The Origin of the Polarization}       
 
The preeminent question raised by this work is the cause of
the distortions that lead to the polarization.  To put this
question in perspective we must examine both the nature of
the polarization and its time dependence.
 
Trammell \etal\ (1993) suggested that the flux 
from a scattering layer above the 
photosphere is already polarized as it impinges on the hydrogen
envelope from below.  
This flux was then envisaged to be absorbed 
upon passage through the hydrogen layer.  
This physical picture, which was implicit in their mode of analysis
and decomposition of the polarized flux into continuum, line,
and ISM effects, was supported by the form of the derived Stokes
flux.  The Stokes flux showed a blue-shifted absorption component
which suggested that the polarization was already imposed on the
continuum at deeper layers and then some of the polarized flux was
scattered or absorbed as it passed through the hydrogen layer.  

An important result of the models presented in this paper is to illustrate 
that the simple idea of a polarized continuum flux impinging on the 
hydrogen layer from below may be both simplistic and incorrect.  The 
reality is more complex.  Trammell \etal\ questioned whether the 
polarization could both be generated within the H layer and still have 
enough optical depth to do the requisite ``absorption"
that is observed in the polarized flux density spectrum.
The current models show that the polarized continuum and the 
depolarization of the line can be generated in the same layer.
The shape of the absorption trough in the polarization 
spectrum, the percent polarization, presented here
is roughly consistent with this.

The interpretation of Trammell \etal\ also has the drawback that if the 
scattering layer is beneath the H envelope, the large intrinsic 
polarization requires that the very large distortions, with axis ratios of 
order 1.5 to 1 or E $\sim$ 0.6 (Shapiro and Sutherland 1982; Trammell 
\etal\ 1993) must be in the helium core.  This level of polarization is 
difficult to generate and maintain in polytropes strongly distorted by 
differential rotation (Steinmetz \& H\"oflich 1992) or by aspherical 
explosions in basically spherical core/envelope configurations because 
sphericity tends to be regained in the expansion.

Another constraint on this interpretation is provided by the
analysis of H\"oflich (1995).  In simple parametrized models
with distorted cores and spherical envelopes, H\"oflich finds
that the polarization is low as long as the spherical envelope
is optically thick. As the photosphere recedes, the polarization
grows to the large values dictated by the distortion of the
core, even suffering a reversal of sign since the long axis
will emerge from the optically thick envelope before the short
axis.  No such strong time-dependence is seen near the
first minimum nor near the second maximum in the light curve,
the two suggested times when the photosphere might recede from 
the envelope to the core.  It thus seems difficult to reconcile
the hypothesis of a distorted core and spherical envelope
with the relatively weak time-dependence of the polarization.

An alternative possibility is that the flux from the core is unpolarized
and that a non-spherical configuration in the envelope creates the
polarization through scattering.  This is the basic ansatz
that is assumed for the models presented in this paper.
A distorted envelope is not out of the  question given the evidence
in many other contexts for bipolar flow
associated with late stages of evolution, especially in binary systems.
An advantage of this picture is that with a
configuration in which an unpolarized, nearly point source (the core)
illuminates a distorted scattering envelope, much less distortion is
required to produce the same degree of polarization (Brown \& McLean 1977;
Fox 1991; H\"oflich 1991;  Trammell \etal\ 1993).  

A distorted envelope 
probably requires some mechanism other than rapid rotation since the
envelope is expected to be extended and convective.  
H\"oflich (1995) has argued that a binary common envelope configuration,
in particular, may naturally yield the necessary distorted envelope.
We also note that with the understanding that the H$\alpha$ profile is
distorted somewhat
by He~I $\lambda$6678, there is no direct evidence in the
spectrum at this epoch for any major asymmetries in the envelope.  As
noted above, however, this does not assure that the envelope is spherically
symmetric.        

H\"oflich (1995) argues that the polarization should be nearly constant 
while the photosphere is in the distorted envelope.  This is consistent at 
some level with the observations from early April to April 20.  The slow 
rise suggested for this interval could be due to the effect of the 
flattening density gradient.  It is not clear, however, that this model can 
account for the observations of Jannuzi \etal\ (1993).  The close agreement 
of the observed size and orientation of the polarization between the 
results of Trammell \etal\ (1993) and of Jannuzi \etal\ gives a reassuring 
self-consistency, but may provide a stringent test of models.

The important difference is that the
observations of Trammell \etal\ on April 20 were very likely
made while the photosphere was still in the hydrogen envelope,
given the lack of evidence for strong lines of He I, as
discussed above.  The observations of Jannuzzi \etal\, were
made only 6 days later, 26 April, but this was a critical week
in the spectral evolution of SN~1993J.  The He I lines,
especially He $\lambda$6678, began to appear between
22 April and 26 April (Hu \etal\ 1993; Swartz \etal\ 1993;
Filippenko, Matheson, \& Ho 1993; Wheeler \& Filippenko 1995).
Thus the observations of Trammell \etal\ were probably made
just before the photosphere receded into the inner helium 
mantle and those of Jannuzi \etal\ just after this transition.
It is thus remarkable that the results of Trammell \etal\
were similar to the polarization determined near the first
minimum and especially
that there was no significant change in the amplitude and 
orientation of the polarization between 20 April
and 26 April.  We do note that 
the small rotation $\sim$ 10 degree 
across H-alpha in the data of Trammell \etal\
becomes a $\sim$ 20 degree rotation in the Steward data, 
even after correction for the interstellar polarization
(Schmidt 1995).
This may be a signature of the photospheric transition,
but it is rather subtle.
It is not clear that the overall constancy of the polarization
amplitude and angle can
be reconciled with notion that the polarization is due to a
distorted hydrogen envelope and that the photosphere receded through
the base of the hydrogen envelope at this time.  
H\"oflich (1995) suggests that envelope 
distortion could be caused by a common
envelope configuration in which the companion was close to
the inner edge of the hydrogen envelope.  It is difficult to
see how the effects of such a companion, 
in the polarimetry or otherwise, would not be seen very 
distinctly at this epoch.

One might argue that the photosphere was sufficiently deep
in the hydrogen envelope even at the time of the Trammell \etal\
observations that there was no change in the polarized component
despite the change in the total spectral features as the 
photosphere receded on into the helium layers.  This is
difficult to reconcile with the fact that the polarization
observed on 4--7 April (corrected for ISP), 
when the photosphere was in the
outer portions of the hydrogen envelope, was essentially
the same as on 20--26 April when the photosphere was
near the base of the hydrogen.  In the former case, the
envelope was still optically thick, in the latter, it
was optically thin;  yet this seems not to have significantly affected
the polarization.
      
H\"oflich (1994) also explores a model in
which the geometry is spherical, but the emitted light
is polarized because the source of the radiation, perhaps
a pulsar or a blob of radioactive nickel, is off-center.  
This model might be consistent with the surprisingly circular
shape of the VLA radio image (Bartel \etal\ 1994)
H\"oflich notes that for consistency of this model
with SN~1993J, the source of the asymmetric flux must have
been at large optical depth even at the end of April, that
is, within layers that were still optically thick even
though the hydrogen layer was rapidly becoming optically
thin.  
A continuum was observed in SN 1993J until at least June 1993,
suggesting that some portions of the ejecta were still
optically thick at that time and that the velocity of the photosphere
was well above 1000 km s$^{-1}$.  Thus an asymmetric energy source
could still be buried for some time.
This picture may be consistent at some level with the 
near constancy of the polarization from early to late April.
Although the polarization varies strongly with optical depth
for moderate to small optical depth (H\"oflich 1990), it varies 
only slowly with optical depth for a buried asymmetric source.
There should be a strong variation of the polarization as any
asymmetric source finally becomes exposed.  The later Lick data
may put important constraints on this hypothesis.  
At the current epoch, two years
later, there is no sign of a pulsar, buried or otherwise,
but asymmetries of the nebular emission lines may be consistent
with a strongly off-center blob of radioactive nickel 
(Chugai and Wang, 1995).
    
We are left with the conclusion that while the current models
have given some important constraints on the size of
the asymmetry and the orientation, we do not have a
clear picture of how the polarization is physically generated.
In particular, it is difficult to see how any model can give
nearly constant polarization in amplitude and orientation,
from early April when the hydrogen-rich envelope was optically thick,
to late April when it was optically thin.  In hindsight,
the observations, sparse as they are, were ideally timed.
The early observations span the first minimum.  If that were
the time when the photosphere receded from the hydrogen to
helium layers then some change in polarization might be expected,
yet none was apparently observed.  The observation in late
April span the time when the spectrum became helium rich,
strongly suggesting that the photosphere receded from the
hydrogen to the helium layers.  The constancy of the polarization
is even more reliable at this epoch.  It is difficult to
see how distorted cores or envelopes could
give this behavior.  An asymmetric source of the flux might be
consistent with the data of the first month.

It is perhaps also worthwhile to consider other means of generating
the observed polarization.  One factor that has not been 
investigated quantitatively here is the circumstellar medium.
We know that such a medium is present from its effects on
the radio and X-ray emission and the continued emission of 
H$\alpha$.  Scattering from an external medium might give the
nearly constant polarization observed, or at least contribute
to the total polarization in a way that has yet not been considered.  
The residual rotation of the polarization at H$\alpha$ after correction
for the ISM in the Arizona data suggests that the line emission
region has a slightly different geometry than the continuum
and hence that there is some factor
involved beyond the ISM and the electron-scattering component 
of the supernova ejecta.
Spectropolarimetry of Nova Cas 93 (Trammell 1995) shows that P-Cygni
line profiles generated in the nova ejecta are scattered and
polarized by an external medium, probably dust rich that is receding
from nova at about 200 km s$^{-1}$.  In this object, at least,
there is clear evidence that the scatterers are located exterior
to the absorption in the lines, in the spirit of the original
suggestion of Trammell \etal\ (1993).  There is the potential for
such an effect in the circumstellar nebula of SN 1993J.  In particular
we note that dust scattering is wavelength independent, like electron
scattering, but of order 100 times more efficient at scattering. 
 
There are, however, several problems with invoking the circumstellar
medium to account for the observations of SN 1993J.  
It would be very difficult for an aspherical
wind to produce a significant total polarization
because of the expected small optical depth to electron scattering,
although dust scattering might be significant if the matter is dusty.
Another important factor is that the broad minimum in the polarization
spectrum presented here must obviously be formed in the supernova
ejecta, not by H$\alpha$ absorption in any external medium.
It may nevertheless be conceivable that an external scattering
screen could be exposed to  relatively more of the continuum emitting volume
than the H$\alpha$ emitting volume due to optical depth effects
within the ejecta.  Whether this sort of configuration could
reproduce the observed polarization as well as the models 
presented here requires quantitative investigation.

To understand the nearly constant polarization and the rotation
of the polarization at H$\alpha$, to check the current models,
and to explore the possible effect of an external dust
scattering medium, it becomes even more imperative to closely
examine the later data obtained at Steward and Lick Observatories
and that published (after this paper was in press) by
Doroshenko, Efimov, and Shakhovskoi (1995).  

The other main question remaining to be answered is whether deviations 
from sphericity are quite common for SN~II or whether SN 1987A and SN 1993J 
were ``bad luck." The answer to this question is critical if SN II are to 
be used as standard candles.  We strongly recommend that polarization 
measurements of supernovae be obtained on a regular basis.

\acknowledgments {PAH takes great pleasure in thanking
Bob Kirshner and all the members of his group for helpful discussions and 
useful comments.  Special thanks go to Al Cameron for his support. 
JCW is grateful to Lifan Wang and Alex Lazarian for reading 
the manuscript and for helpful comments.  This 
work has been supported by the Deutsche Forschungsgemeinschaft through the 
grant Ho~1177/2-1, by NSF Grant AST 9218035 and by NASA grant
NAGW 2905. 
Some of the computations were
performed at NCSA under the grant AST940002N.}

%


\newpage           
     
\begin{thebibliography}
\overfullrule=0pt
\bibitem[Abbot and Lucy 1985]{Abbot-Lucy-85}
\reference
Abbot, D. C., \& Lucy  L. B. 1985, ApJ, 288, 679

\bibitem[Arnett \etal\ 1989]{Arnett-et-al-89}
\reference
Arnett, W. D., Bahcall, J. N., Kirshner, R. P., \& Woosley, S. E.
1989, ARA\&R, 27, 629
     
\bibitem[Baade 1926]{Baade-26}
\reference Baade, W. 1926, Astron. Nachrichten, 228, 359

\bibitem[Baron \etal\ 1993]{Baron-et-al-93}
\reference
Baron, E., Hauschildt, P. H., Branch, D., Wagner, R. M., Austin, S. J., 
Filippenko,  A. V., \& Matheson, T. 1993, ApJ, 416, L21 
 
\bibitem[Baron \etal\ 1994]{Baron-et-al-94}
\reference
Baron, E., Hauschildt, P. H.,\& Branch D. 1994, ApJ, 426, 334  

\bibitem[Baron \etal\ 1995]{Baron-et-al-95}
\reference
Baron, E., Hauschildt, P., \& Young, T. R. 1995, Phys. Reports, 256, 33

\bibitem[Bartel \etal\ 1994]{Bartel-94}
\reference
Bartel, N., \etal\ 1994, Nature, 368, 610   

\bibitem[Baschek \etal\ 1991]{Baschek-91}
\reference
Baschek, R., Scholz, M., \& Wehrse, R. 1991, A\&A, 246, 374

\bibitem[Branch \etal\ 1981]{Branch-et-al-81}
\reference
Branch, D., Falk S., McCall, M., Rybicki, P., Uomoto, A.K., 
\& Wills, B.J. 1981, ApJ, 244,780
 
\bibitem[Bjorkman \etal\ 1993]{Bjorkman-et-al-93}
\reference
 Bjorkman,  K., \& Nordsiek, K. 1993, private communication with
Wheeler, see also Wheeler \& Filippenko (1995)
 
\bibitem[Brown and McLean 1977]{Brown-McLean-77}
\reference
Brown, J. C., McLean, I. S. 1977, A\&A, 57, 141

\bibitem[Wheeler \etal\  1994]{Wheeler-et-al-94}
\reference
Chugai, N. \& Wang, L. 1995, in preparation
 
\bibitem[Clochiatti \etal\ 1988]{Clochiatti-et-al-88}
\reference
Clocchiatti, A., M\'endez, M., Benvenuto, O., Feinstein, C., Marraco, H.,
Garc\'{\i}a, B., \& Morrell, N., 1988, Proc. of the 4th George Mason
Conference on SN 1987A, ed. M. Kafatos and A. Michalitsianos, Cambridge
University Press, 70

\bibitem[Shapiro \& Sutherland 1982]{Shapiro-Sutherland-82}
\reference
Clocchiatti, A., Wheeler, J. C., Barker, E. S., Filippenko, A. V., 
Matheson, T., \& Leibert, J. W. 1995, ApJ, 446, 167

\bibitem[Filippenko \& Wheeler 1995]{Filippenko-Wheeler-95}         
\reference
Cropper, M. S., \etal\ 1987,  MNRAS, 231, 695

\bibitem[Daniel 1982]{Daniel-82}                                   
\reference
Daniel, J. Y. 1982, AAP, 86, 198

\bibitem[Shapiro \& Sutherland 1982]{Shapiro-Sutherland-82}
\reference
Doroshenko, V. T., Efimov, Yu. S., \& Shakhovskoi, N. M. 1995, 
Astron. Lett., 21, 513

\bibitem[Duschinger \etal\ 1994]{Duschinger-et-al-94}              
\reference
Duschinger, M., Branch, D., H\"oflich P., Kudritzki R. P., \& Puls 1994, 
AAP, in press  
 
\bibitem[Eastman \& Kirshner 1989]{Eastman-89}                           
\reference
Eastman, R., Kirshner, P. R. 1989, ApJ, 347, 771

\bibitem[Wheeler \etal\  1994]{Wheeler-et-al-94}
\reference
Filippenko, A. V., Matheson, T., \& Ho, L. C. 1993, ApJ, 415, L103 
 
\bibitem[Eastman \& Kirshner 1989]{Eastman-89}                           
\reference
Fox, G. K. 1991, ApJ, 379, 663 
 
\bibitem[Harkness 1987, 1991]{Harkness-91}         
\reference
Harkness, R .P.  1991,  SN 1987A and Other Supernovae, ed. J. Danziger
\& K. Kj\"ar, ESO, Garching, 447
 
\bibitem[Hershkowitz \etal\ 1986]{Hershkowitz-et-al-86}
\reference
 Hershkowitz, S., Lindner, E., \& Wagoner, R. V. 1986, AAP, 301, 220
 
\bibitem[Hillebrandt \& H\"oflich 1990]{Hillebrandt-et-al-90}
\reference
Hillebrandt, W., H\"oflich, P. 1990, Reports on Prog. in Phys., 52, 1421
 
\bibitem[H\"oflich 1988]{Hoeflich-88}                  
\reference
H\"oflich, P. 1988, Proc. Astron. Soc. Austrl., 7, 434
 
\bibitem[H\"oflich \etal\ 1989]{Hoeflich-et-al-86}                  
\reference
H\"oflich,  P. 1989, in Particle Astrophysics, 
ed. E.B. Norman, World Scientific, Singapore, 205

\bibitem[H\"oflich 1990]{Hoeflich-90}                  
\reference
H\"oflich,  P. 1990,  Analysis of type II supernovae at the photospheric
phase, Munich, habilitation thesis, MPA 563
 
\bibitem[H\"oflich 1991]{Hoeflich-91}                  
\reference
 H\"oflich, P. 1991, AAP, 246, 481
 
\bibitem[H\"oflich 1991]{Hoeflich-91}                  
\reference
 H\"oflich, P. 1991b, in Supernovae, ed. S.E. Woosley 
 (Springer Press: New York), 415

\bibitem[H\"oflich 1994b]{Hoeflich-94b}                  
\reference
 H\"oflich, P. 1995, ApJ, 443, 89

\bibitem[H\"oflich \etal\ 1993]{Hoeflich-et-al-93}                  
\reference
H\"oflich P., Langer N., \& Duschinger, M. 1993, AAP, 275, L29
 
\bibitem[H\"oflich \& Steinmetz 1991]{Hoeflich-Steinmetz-91}                  
 \reference
 H\"oflich, P., \& Steinmetz, M. 1991, Nuclear Astrophysics, P5, ed.
W. Hillebrandt \& E. M\"uller, 103
 
\bibitem[H\"oflich \etal\ 1986]{Hoeflich-et-al-86}                  
\reference
H\"oflich,  P., Wehrse, R., Shaviv, G. 1986, AAP, 163, 105
 
\bibitem[Duschinger \etal\ 1994]{Duschinger-et-al-94}              
\reference
Hu, J. Y., Li, Z. W., Jiang, X. J., \& Wang, I. F. 1993, IAU Circ. No. 5777

\bibitem[Daniel 1982]{Daniel-82}                                   
\reference
Januzzi, B., Schmidt, G. D., Elston, R. E., \& Smith 1993, IAU Circ. No. 5776

\bibitem[Jeffery 1987]{Jeffery-87}                          
\reference
Jeffery,  D.J. 1987, Nature, 329, 419
 
\bibitem[Jeffery 1991]{Jeffery-91}                          
\reference
Jeffery,  D.J. 1991, ApJ, 375, 264
 
\bibitem[Jeffery \etal\ 1993]{Jeffery-et-al-93}                          
\reference
Jeffery, D. J., Kirshner, R. P., Challis, P. M., Pun, C.S.J, 
Filippenko, A. V., Matheson, T., Branch,  D., Chevalier, R. A., 
Fransson, C., Panagia, N., Wagoner,  R. V., Wheeler, J. C., 
\& Clocciatti,  A. 1994, ApJ, 421, L27

\bibitem[Daniel 1982]{Daniel-82}                                   
\reference
Kraus, M., Beck, R., Hummel, E., 1989, AAP, 217, 17

\bibitem[Lucy 1988] {Lucy-88}                          
\reference
Lucy, L. B. 1988, Proc. of the 4th George Mason conference, 
ed. M. Kafatos, Cambridge University Press, 323
 
\bibitem[Lundqvist \etal\ 1993]{Lundqvist-et-al-93}    
\reference
Fransson, C., Lundqvist, P., \& Chevalier, R. A. 1994, ApJ, in preparation.

\bibitem[Duschinger \etal\ 1994]{Duschinger-et-al-94}              
\reference
McCall, M. L. 1985, in Supernova as Distance Indicators, ed. N. Bartel,
(Berlin: Springer-Verlag), 48 

\bibitem[Duschinger \etal\ 1994]{Duschinger-et-al-94}              
\reference
McCall, M. L., Reid, N., Bessell, M. S., \& Wickramasinghe, D. 1984, 
MNRAS, 210, 839

\bibitem[M\'endez \etal\ 1993]{Mendez-et-al-88}
\reference
M\'endez, M., Clocchiatti, A., Benvenuto,  O. G., Feinstein,  C., 
Marraco,  H. G., 1988, ApJ, 334, 295
 
\bibitem[Mihalas 1978] {Mihalas-78}
\reference
Mihalas, D. 1978, Stellar Atmospheres, W.H. Freeman \& Company, San
Francisco, 2nd edition
 
\bibitem[Lundqvist \etal\ 1993]{Lundqvist-et-al-93}    
\reference 
Miller, J. private communication, 1993

\bibitem[Mihalas 1978] {Mihalas-78}
\reference
Miller, J. S. \& Goodrich, R. W. 1990 ApJ, 355, 456

\bibitem[Nomoto \etal\ 1993]{Nomoto-et-al-93}                
\reference
 Nomoto K., Susuki T., Shigeyama T., Kumagai S., Yamaoka, H., Saio H.
1993, Nature,  327, 287

\bibitem[M\'endez \etal\ 1993]{Mendez-et-al-88}
\reference
\noindent Schmidt, B., Kirshner, R. P., Eastman, R. G., 
Phillips, M. M., Suntzeff, N. B., Hamuy, M., Maza, J., \&
Aviles, R. 1994, ApJ, 432, 42 
 
\bibitem[Schmid-Burgk 1982]{Schmid-Burgk-82}                         
\reference
Schmid-Burgk, J. 1982, AAP, 108, 169

\bibitem[Shapiro \& Sutherland 1982]{Shapiro-Sutherland-82}
\reference
Schmidt, G. 1995, private communication

\bibitem[Shapiro \& Sutherland 1982]{Shapiro-Sutherland-82}
\reference
Serkowski, K. 1970, ApJ, 160, 1083

\bibitem[Shapiro \& Sutherland 1982]{Shapiro-Sutherland-82}
\reference
Shakhovskoi, N. M., and Efimov, Yu. S. 1973, Sov. Astron. AJ, 16, 7

\bibitem[Shapiro \& Sutherland 1982]{Shapiro-Sutherland-82}
\reference
Shapiro P.R., Sutherland P.G. 1982, ApJ, 263, 902
 
\bibitem[Shapiro \& Sutherland 1982]{Shapiro-Sutherland-82}
\reference
Shigeyama, T., Suzuki, T., Kumugai, S., Nomoto, K., 
Saio, H., \& Yamaoka, H. 1994, ApJ, 420, 341

\bibitem[Simmons \& Stewart 1985]{Simmons-Stewart-85}
\reference
Simmons, J.F.L., \& Stewart, B. G., 1985, AA, 142, 100

\bibitem[Smith 1993]{Smith-1993}
\reference
 Smith, P. 1993, private communication

\bibitem[Shapiro \& Sutherland 1982]{Shapiro-Sutherland-82}
\reference
Spyromilio, J., \& Bailey, J. 1993,  Proc.  ASA, 10, 263

\bibitem[Sobolev 1957]{Sobolev-57}                                     
\reference
Sobolev, V. V., 1957, Sov. Astron. J., 1, 297
 
\bibitem[Steinmetz \& H\"oflich 1992]{Steinmetz-Hoeflich-92}
\reference
Steinmetz,  M., \&  H\"oflich, P. 1992, AAP, 257, 641
 
\bibitem[M\'endez \etal\ 1993]{Mendez-et-al-88}
\reference
Swartz, D. A., Clocchiatti, A., Benjamin, R., 
Lester, D. F., \& Wheeler, J. C.  1993, Nature, 365, 232
 
\bibitem[Trammell \etal\ 1993]{Trammell-et-al-93}
\reference
 Trammell,  S. R., Hines, D. C., \& Wheeler, J. C. 1993, ApJ, 414, L21

\bibitem[Nomoto \etal\ 1993]{Nomoto-et-al-93}                
\reference
Tran, H. \& Filippenko, A. V. 1993, private communication

 \bibitem[Wardel \& Kronberg 1974]{Wardle-Kron-74}
\reference
Utrobin, V. P. 1995, AAP, in press.
 
\bibitem[Filippenko \& Wheeler 1995]{Filippenko-Wheeler-95}         
\reference
Van Dyk, S., Weiler, K. W., Sramek, R. A., Rupen, M. P., 
\& Panagia, N. 1994, ApJ, 432, L115

\bibitem[Shapiro \& Sutherland 1982]{Shapiro-Sutherland-82}
\reference
Wang, L., Li, Z., Clocchiatti, A., Wheeler, J. C. \& Wills, D.
1995, submitted to ApJ 

 \bibitem[Wardel \& Kronberg 1974]{Wardle-Kron-74}
\reference
Wardle, J.F.C., \& Kronberg, P. P. 1974 , ApJ, 194, 249

\bibitem[Filippenko \& Wheeler 1995]{Filippenko-Wheeler-95}         
\reference
Wheeler, J. C. \& Filippenko, A. V. 1995, Review of the Contributions
to the Xian Workshop on SN 1993J, ed. D. McCray, 
(Cambridge: Cambridge University Press), in press

\bibitem[Shapiro \& Sutherland 1982]{Shapiro-Sutherland-82}
\reference
Wheeler, J. C., Barker, E., Benjamin, R.,
Boisseau, J., Clocchiatti, A., de Vaucouleurs, G., 
Gaffney, N., Harkness, R. P., Khokhlov, A. M., Lester, D. F.,
Smith, B. J., Smith, V. V., \& Tomkin, J. 1993, ApJ, 417, L71
 
\bibitem[Wheeler \etal\  1994]{Wheeler-et-al-94}
Wheeler, J.C., Harkness, R.P., Clocchiatti, A., Benetti, S., Brotherton, M.,
Depoy, D., Elias, J. 1994, ApJ, 436, L135

\bibitem[Shapiro \& Sutherland 1982]{Shapiro-Sutherland-82}
\reference
Wilking, B. A., Lebofsky, M. J., \& Rieke, G. H., 1982, AJ, 87, 695
 
\bibitem[Wesselink 1946]{Wesselink-46}
\reference
Wesselink, A.J., 1946,  Bull. Astron. Inst. Neth., 368, 91

\bibitem[Sobolev 1957]{Sobolev-57}                                     
\reference
Wolstencroft, R. D., \& Kemp, J. C. 1972, Nature, 238, 45

\bibitem[Steinmetz \& H\"oflich 1992]{Steinmetz-Hoeflich-92}
\reference
Woosley, S. E., Eastman, R. G., Weaver, T. A., \&  Pinto 1994,
ApJ, 429, 300
 
\bibitem[Yamada \& Sato 1991]{Yamada-Sato-91}
\reference
 Yamada, S., Sato, K. 1991  ApJ, 358, L9
\end{thebibliography}
\end{document}